\documentclass[fleqn,usenatbib]{mnras}
\usepackage[T1]{fontenc}
\usepackage{ae,aecompl}
\usepackage{graphicx}   
\usepackage{amsmath}    
\usepackage{amssymb}    
\usepackage{float}
\usepackage[normalem]{ulem}

\usepackage[dvipsnames]{xcolor}
\usepackage{orcidlink}

\title[Jet MHD analytical model]%
{Jet MHD analytical model --- framework for radiative transfer}
\author[Nokhrina]{\parbox{\textwidth}{
E.~E.~Nokhrina$^{1,2}$\thanks{E-mail: nokhrinaelena@gmail.com}, I.~N.~Pashchenko$^{3}$\orcidlink{0000-0002-9404-7023}, V.~A.~Frolova$^{2}$, R.V. Todorov$^{3}$
}
\vspace{0.4cm}\\
\parbox{\textwidth}{
$^1$Division of Theoretical Physics, P.N. Lebedev Physical Institute, Leninsky prosp.~53, Moscow, 119991, Russia \\
$^2$Moscow Center for Advanced Studies, Kulakova str. 20, Moscow, 123592, Russia \\
$^3$Astro Space Center, P.N. Lebedev Physical Institute, Leninsky prosp.~53, Moscow, 119991, Russia
}
}

\begin{document}

\date{Accepted 2005 October 3. Received 2005 October 3; in original form 2025 July 14}

\pagerange{\pageref{firstpage}--\pageref{lastpage}} \pubyear{}

\maketitle

\label{firstpage}

\begin{abstract}
We developed the full magnetohydrodynamical analytical jet model that allows accurate reproducing of a transversal and longitudinal structure for a highly collimated relativistic jets. This model can be used as a setup for convenient solution of radiative transfer equations and modelling the total intensity and polarization maps.  
We show that the analytical fits are in excellent agreement with the numerical solutions of full magnetohydrodynamical equations. Our approach allows setting easily different models for an emitting plasma number density. For example, we show that the equipartition number density ranges from several to tens of percent of a total number density. We show that the Doppler-corrected emissivity distribution behave in such a way that we may expect a limb brightened intensity pattern on a sub-parsec scale and a spine-brightened structure downstream. We reproduce the broken power-law dependence of a jet pressure at its boundary from the jet radius. The corresponding power exponents are in agreement with the parabola-to-cone transition observed directly in nearby sources.
\end{abstract}


\section{Introduction}
\label{s:intro}

Relativistic jets in active galactic nuclei (AGN) are a phenomenon with physics, seemingly understood in general, but still posing many questions in important details. The central engine of an AGN consists of a rotating supermassive black hole \citep[see e.g.][and references within]{Beskin10, BMR-19}, accreting matter in different states \citep{NIA03, Tchekhovskoy_11, McKinney12, ZCST14}. The collimation of jets is attributed to the ambient medium \citep[see e.g.][]{Lyu09, Asada12, BCKN-17, Kov-20, NKP20_r2}. The initially magnetically dominated outflow transfers its energy flux into the bulk plasma motion, with the observed Lorentz factor\sout{s} values reaching $50-100$ \citep{MOJAVE_XVII}. Among the main powering mechanisms for a jet are considered a Blandford--Znajek process that extracts a central supermassive black hole rotational energy \citep{BZ-77} and/or a Blandford--Payne process with a jet launched by an accretion disc \citep{BP-82}. These models provide different scenarios for possible magnetic field, plasma number density and bulk velocity distributions. As a jet emission in radio and millimetre wavelengths is well described by the synchrotron self-absorbed emission of relativistic plasma \citep{BlandfordKoenigl1979}, the particular launching and plasma acceleration mechanism affects the observed emission in total and polarized spectral flux. 

Observations with Very Long Baseline Interferometry (VLBI)
reveal diverse patterns of intensity and polarization maps of jets from AGNs. The most striking phenomenon observed in total intensity is a limb brightening. In some sources limb brightened jets are detected only near the jet base, while in the other sources up to much larger distances. Bright ridges have been observed on the scales of up to 2--4 milliseconds of arc (projected) for nearby sources: 3C~84 \citep{Nagai14, Giov18}, Mrk~510 \citep{Piner09, Koyama19}, Cyg~A \citep{Boccardi16}, 3C~264 \citep{Boccardi19}, NGC~315 \citep{Park24} and Cen~A \citep{Janssen21}. The jet in M~87 demonstrates a triple brightened structure at its base \citep{Walker18, Lu23} and double to triple brightened structure at much larger distances along a jet \citep{Nikonov23}. The central bright ridge may be the artefact of the specific data processing \citep{Pashchenko23}, and two other ridges may be explained by the developed Kelvin--Helmholtz instability \citep{Nikonov23}. The jet in 3C~273 is characterized by limb brightening up to the distance more than 20~mas \citep{Bruni2021}.

Explaining the observed diverse phenomena needs the appropriate jet modelling. As the total and polarized synchrotron emission distribution, including limb brightening, depends on the local values of an emitting plasma number density $n$, magnetic field $B$ and its orientation, emitting plasma pitch-angle (angle between velocity and magnetic field in a plasma proper frame) distribution function and plasma bulk velocity \citep[see, e.g.][]{GinzburgSyrovatskii65,Lyutikov2005}. Thus, one needs two principal constituents to reproduce a jet spectral intensity and polarization distribution: the underlying jet structure and the spatial and energy distribution of emitting plasma. The latter, in general, does not necessarily depend on the former.

The problem of limb brightening has been addressed in many works. The velocity stratification may account for this effect. If a jet consists of a faster spine with Lorentz Factor $\Gamma_\mathrm{sp}$ and slower sheath with $\Gamma_\mathrm{sh}$, for the viewing angles $\Gamma_\mathrm{sp}^{-1}<\theta_\mathrm{v}<\Gamma_\mathrm{sh}^{-1}$ jet emission is suppressed for higher velocity and boosted for slower one \citep{Nagai14, Walker18, Giov18, Boccardi16, Janssen21}. This explanation does not take into consideration the possibility of an emissivity stratification. If the magnetic field and/or emitting plasma number density decreases towards the jet edge, leading to a drop in a plasma emissivity, this effect may compensate or even prevail over the relativistic boosting effect.

The effects of a prescribed magnetic field and its topology distribution across a jet on a total intensity and polarization maps have been explored by \citet{KramerMacDonald2021}. They conducted 3D relativistic magnetohydrodynamic jet simulations, in which they probed predominantly poloidal, helical and toroidal magnetic field and emitting particles distributions proportional to thermal, internal and magnetic energy densities. They concluded that edge-brightened and spine-brightened emission favour toroidal and poloidal magnetic field correspondingly, while the effect of emitting distribution choice is insignificant.    

The emission, calculated for a numerically modelled within general relativity magnetohydrodynamics jet \citep{Fromm2022, CruzOsorio2022}, demonstrates the high impact of the spine with the variation of $\sigma_{\rm cut}$ --- the threshold value of magnetization which defines the separation between the sheath and spine, and the limb brightening was obtained for exclusion of central axial regions with a high magnetization, i.e. the spine. In this work, the emitting distribution combined thermal and non-thermal electrons.

The limb brightening has been obtained for a magnetically dominated part of a jet in the force--free approximation, in which the plasma inertia is neglected. The magnetic field configuration for such models is proposed by \citet{BroderickLoeb2009}. \citet{Takahashi2018} and \citet{Ogihara19} used the ring-like spatial distribution of emitting plasma in order to obtain the limb brightening and to explore how the black hole spin affects the symmetry of the obtained bright ridges. The prescription of a plasma density proportional to the Poynting flux at the jet base, growing towards the boundary, allowed \citet{Hir25} to reproduce both limb brightening at the jet base and an observed in M~87 on scales of several gravitational radii ring-like structure \citep{Lu23}. 

Downstream, where the Poynting flux does not strongly dominate the total energy density flux --- starting from a few to ten parsecs along a jet \citep{Kov-20}, --- the full magnetohydrodynamics (MHD) approach must be used. \citet{Frolova23} used the numerical solution of MHD equations for the models with a constant angular velocity \citep{Lyu09} and a closed electric current \citep{BCKN-17} to show the dominance in emission of a central core \citep{Kom07, Beskin09, Lyu09} and a need to suppress the emissivity of this core in order to account for the detected jet emission spectral flux. However, solving the radiative transfer equations using the numerically obtained jet structure is both time- and computational resources- consuming. This especially holds if the inference of the jet physical parameters from the observed data \citep{Porth11,2018Galax...6...31A} is made by the means of a computationally expensive numerical methods, e.g Bayesian methods like Markov Chain Monte Carlo \citep[e.g.][]{2019MNRAS.488..939P} or optimization routines \citep[e.g.][]{2019A&A...629A...4F}.

The purpose of this paper is to find an analytical approximation to the numerical solution of full MHD equations.
The obtained within this approach magnetic field, plasma bulk motion velocity and total number density distributions determine synchrotron transfer coefficients, and can be easily implemented into the radiative transfer equations code.
Such an analytical model provides a convenient and flexible instrument to model the jet structure to probe how different emitting plasma distributions affect the total and polarized intensity maps and allows for precise control of jet parameters. The model proposed here is based on the paper by \citet{Beskin24}, therefore below we refer the work \citet{Beskin24} as B24.

The paper is organised as follows. In section~\ref{s:sec1} we develop the analytical approximation for electric and magnetic fields reproducing the results of the full magnetohydrodynamics equations numerical solution in a one-dimensional approach for a highly collimated outflows. We present the model profiles for a plasma bulk motion velocity and Lorentz factor in section~\ref{s:vel}, for a total and equipartition plasma number density in section~\ref{s:n}. We discuss the jet pressure and obtain a universal broken power-law fit for $P_\mathrm{jet}(r_\mathrm{jet})$ in section~\ref{s:pressure}. In section~\ref{s:intensity} we probe some illustrations of the total intensity profile across a jet for the model. 
Finally, we discuss the results and possible applications of a method and summarise our work in section~\ref{s:Conc}.

\section{Electromagnetic field model}
\label{s:sec1}

We use the model introduced by \citet{Beskin24} to describe the electric and magnetic field in a jet.
Below the cylindrical coordinates $\{r;\,\varphi;\,z\}$ are used to map the jet profile. A magnetic flux is a function $\Psi(r,\,z)$ of $r$- and $z$-coordinates. We assume the following definitions for magnetic and electric field components:
\begin{equation}
{\bf{B}}_\mathrm{p}=\frac{\nabla\Psi\times e_{\varphi}}{2\pi r},
\label{Bp0}
\end{equation}
\begin{equation}
B_{\varphi}=-(1+\varepsilon)\frac{\Omega_\mathrm{F}}{2\pi c}\left|\nabla\Psi\right|,
\label{Bphi0}
\end{equation}
\begin{equation}
{\bf{E}}=-\frac{\Omega_\mathrm{F}}{2\pi c}\nabla\Psi.
\label{E0}
\end{equation}
The angular velocity $\Omega_\mathrm{F}$ is a function of $\Psi$ and must be defined depending on a particular model (see details in B24).
The Equations~(\ref{Bp0}) and (\ref{E0}) are the exact definitions adopted in magnetohydrodynamics. The Equation~(\ref{Bphi0}) is written within the assumption of a relativistic outflow, where a toroidal magnetic field exceeds slightly an electric field. The difference between $B_{\varphi}$ and $E$ is parametrised by the value $\varepsilon$ and defines the Lorentz factor across a jet (see details in Section~\ref{s:vel}). Parameter $\varepsilon$ can be constant, as in B24, or non-constant, as in this work.

The jet has a different structure depending on the magnitude of a poloidal component at the axis $B_0$. The critical value for $B_0$ is
\begin{equation}
B_\mathrm{cr}=\frac{\Psi_0}{\pi R_\mathrm{L}^2\Gamma_\mathrm{in}\sigma_\mathrm{M}}=\frac{B_\mathrm{L}}{\Gamma_\mathrm{in}\sigma_\mathrm{M}},
\label{Bcr}
\end{equation}
where $\Psi_0$ is a total magnetic flux in a jet, $R_\mathrm{L}=c/\Omega_0$ is a light cylinder radius defined by the typical (or constant) angular velocity $\Omega_\mathrm{F}=\Omega_0$, $\Gamma_\mathrm{in}$ is an injection Lorentz factor of plasma near the axis. It is convenient to introduce the characteristic magnetic field $B_\mathrm{L}=\Psi_0/\pi R_\mathrm{L}^2$. The Michel's magnetization parameter $\sigma_\mathrm{M}$ is equal to the ratio of Poynting to the plasma kinetic energy flux at the jet base
\begin{equation}
\sigma_\mathrm{M}=\frac{\Omega_0^2\Psi_0}{8\pi^2 m_\mathrm{e}c^4\eta}
\label{sigmaM}
\end{equation}
and it defines the maximum bulk motion Lorentz factor $\Gamma_\mathrm{max}=\Gamma_\mathrm{in}+2\sigma_\mathrm{M}$ for $\Omega_\mathrm{F}=\mathrm{const}$ \citep{Nokhrina2022,Beskin2025}. Mass-to-magnetic flux function $\eta(\Psi)$ defines the ratio of a poloidal plasma number flux to a magnetic field flux by the Equation~(\ref{n0}), and in this work it is assumed constant.
Poloidal magnetic field at the jet axis $B_0$ depends on a distance along a jet $z$ and decreases as the jet propagates and gets wider. Solution of a set of MHD equations in the cylindrical approach shows the appearance of a jet central core (CC) --- the innermost part of a jet having a radius of the order of a light cylinder radius with a constant poloidal magnetic field $B_0$ \citep{Kom09, Beskin09}. The critical magnetic field $B_\mathrm{cr}$ defines a minimum value of $B_0$ needed for the total magnetic flux of a jet to be contained within the central core. Thus, up to some distance along a jet, where the condition $B_0>B_\mathrm{cr}$ still holds, the poloidal magnetic field across the whole jet remains uniform, and we call such part of a jet a central core-dominated flow (CC-dominated flow). For $B_0<B_\mathrm{cr}$ the central core contains only a part of a total magnetic flux. The poloidal magnetic field in this region can be described by the presence of a CC, where it is uniform, and the outer part, where it decreases as a power law.
 
Thus, for $B_0>B_\mathrm{cr}$, the central core dominates a jet of a radius
$r_\mathrm{jet}$ \citep{Beskin09, Beskin24}.
In this case, the magnetic flux is given by
\begin{equation}
\Psi(r)=\Psi_0\left(\frac{r}{r_\mathrm{jet}}\right)^2.
\label{Psi_stream}
\end{equation}
The definitions (\ref{Bp0})--(\ref{E0}) provide
\begin{equation}
B_\mathrm{p}=B_0=\frac{\Psi_0}{\pi r_\mathrm{jet}^2},
\label{Bp1}
\end{equation}
\begin{equation}
B_{\varphi}=-(1+\varepsilon)B_\mathrm{p}\frac{r}{R_\mathrm{L}},
\label{Bphi1}
\end{equation}
\begin{equation}
E=-B_\mathrm{p}\frac{r}{R_\mathrm{L}}.
\label{E1}
\end{equation}
From Equations~(\ref{Bcr}) and (\ref{Bp1}) it follows that a jet sustains the central core-dominated structure up to reaching the radius $r_\mathrm{jet}\approx R_\mathrm{L}\sqrt{\Gamma_\mathrm{in}\sigma_\mathrm{M}}$ in full agreement with the work by \citet{Beskin09}.

For $B_0<B_\mathrm{cr}$ we introduce the magnetic flux function at a given distance $z$ from the jet base as (B24):
\begin{equation}
\Psi(r,\;z)=\frac{\pi B_0(z)r_\mathrm{core}^2}{1-\alpha(z)/2}\left[\left(1+\frac{r^2}{r_\mathrm{core}^2}\right)^{1-\alpha(z)/2}-1\right].
\label{Psi_core}
\end{equation}
Here $r_\mathrm{core}\approx$ a few $R_\mathrm{L}$, as analytical and numerical modelling shows. In this case the dense core with a radius $r_\mathrm{core}$ appears in a jet with the poloidal magnetic field amplitude
\begin{equation}
B_\mathrm{p}=B_0(z)\left[1+\left(\frac{r}{r_\mathrm{core}}\right)^2\right]^{-\alpha(z)/2}.
\label{Bp2}
\end{equation}
The toroidal magnetic and electric fields are given by Equations~(\ref{Bphi1}) and (\ref{E1}).

The index of a poloidal magnetic field decrease with a radius $\alpha(z)$ and a magnetic field at the jet axis $B_0(z)$ define the electromagnetic fields for the whole jet. It is more convenient to use the jet radius $r_\mathrm{jet}$ as an argument instead of $z$. One can set directly the jet boundary shape as an explicit function $r_\mathrm{jet}(z)$. It is also possible to calculate the function $r_\mathrm{jet}(z)$ implicitly by assuming an equilibrium of a jet pressure $P(r_\mathrm{jet})$ and an ambient medium pressure $P_\mathrm{ext}(z)$ (see Section~\ref{s:pressure}). Below we use the designation $B_0(r_\mathrm{jet})$ for a function $B_0(r_\mathrm{jet}(z))$.

Our aim is to find an approximation for $\alpha(r_\mathrm{jet})$. For a known $\alpha$ we define the magnitude of a poloidal magnetic field at an axis $B_0(r_\mathrm{jet})$ using a conservation of a total magnetic flux (\ref{Psi_core}) (see also Equation~(8) by B24):
\begin{equation}
B_0(r_\mathrm{jet})=\frac{\Psi_0}{\pi r_\mathrm{core}^2}\frac{(1-\alpha(r_\mathrm{jet})/2)}{\left(1+r_\mathrm{jet}^2/r_\mathrm{core}^2\right)^{1-\alpha(r_\mathrm{jet})/2}-1}.
\label{B0_core}
\end{equation}

\begin{figure}
    \centering
    \includegraphics[width=1.0\linewidth]{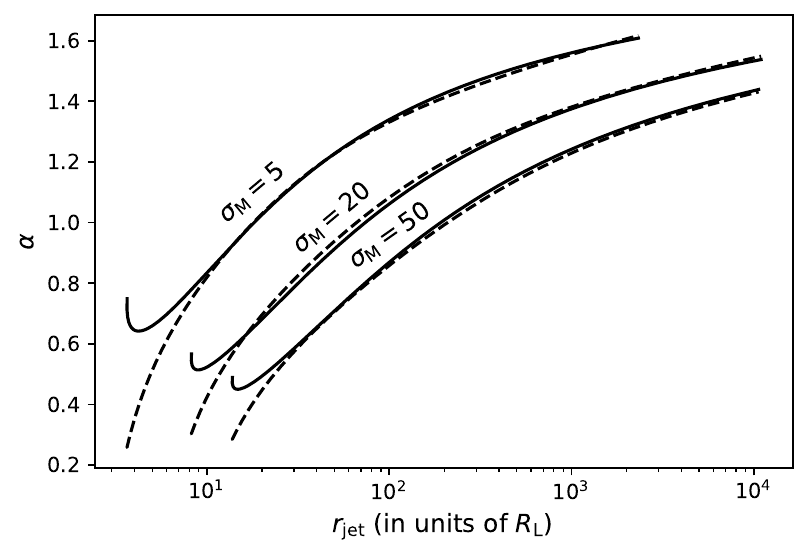} 
    \caption{Solid lines --- $\alpha$, calculated numerically to match the jet pressure at its boundary. Dashed lines --- fits using Equation~(\ref{alpha1}).
    }
    \label{f:alpha}
\end{figure}

To find $\alpha(r_\mathrm{jet})$ we use the analytical solution of cylindrical MHD Equations~(58)--(59) with the integrals~(52)--(55) in \citet{Beskin06} and fit the $B_\mathrm{p}$ with a core approximation (\ref{Bp2}) so that both values are equal at the jet boundary. Here we set $r_\mathrm{core}=R_\mathrm{L}$. The resultant function $\alpha(r_\mathrm{jet})$ is presented in Figure~\ref{f:alpha} in solid lines for an initial magnetization $\sigma_\mathrm{M}=5,\;20,\;50$, corresponding to maximum bulk motion Lorentz factor $\Gamma_\mathrm{max}=12,\;42,\;102$, given by Equation~(A9) in \citet{Nokhrina2022}. The non-monotonous behaviour at small distances $x$ is due to the poloidal magnetic field profile across a jet, although having an almost constant value, deviating from the core-like structure given by Equation~(\ref{Bp2}): see Figure~\ref{f:Bp}, upper left panel. 
That is why we extrapolate values for a parameter $\alpha$ at small jet radii so as to fulfil the condition $\alpha(\sqrt{\Gamma_\mathrm{in}\sigma_\mathrm{M}},\;\sigma_\mathrm{M})=0$, corresponding to Equation~(\ref{Bp1}). Fits (\ref{alpha1}) and (\ref{params}) for $\alpha$ are shown in Figure~\ref{f:alpha} in dashed lines.
The following fit is a good universal approximation for different values of $\sigma_\mathrm{M}$ $\alpha(x_\mathrm{jet})$:
\begin{equation}
\begin{array}{l}
\displaystyle\alpha(x_\mathrm{jet},\,\sigma_\mathrm{M})=1.3\left[\frac{\ln\left(x_\mathrm{jet}-\sqrt{\Gamma_\mathrm{in}\sigma_{\mathrm{M}}}+1\right)}{c_0}\right]^{a}\times \\ \ \\
\displaystyle \times\left[1+\left[\frac{\ln\left(x_\mathrm{jet}-\sqrt{\Gamma_\mathrm{in}\sigma_{\mathrm{M}}}+1\right)}{c_0}\right]^6\right]^{(0.33-a)/6},
\end{array}
\label{alpha1}
\end{equation}
where coefficient $c_0$ and exponent $a$ are the following functions of the initial magnetization:
\begin{equation}
c_0=2.8\,\sigma_\mathrm{M}^{0.22},
\quad
a=0.52\,\sigma_\mathrm{M}^{0.18}.
\label{params}
\end{equation} 

In Figures~\ref{f:Bp} and \ref{f:Bphi} we compare the numerical solution of trans-field equations with our analytical approximations (\ref{Bphi1})--(\ref{Bp2}) and (\ref{alpha1})--(\ref{params}). For each physical value we reproduce its behaviour as a function of a radial distance in units of a light cylinder radius $x=r/R_\mathrm{L}$ for four different distances along a jet, characterised by consequently increasing radius (the jet radius $x_\mathrm{jet}$ grows from upper left to lower right panels). Dotted lines represent numerical solution of MHD equations in cylindrical approximation. Dashed lines show the analytical fits for $B_0(z)$ taken from numerical simulations. Solid lines show analytical approximation with $B_0(z)$ calculated using the total magnetic flux conservation in Equation~({\ref{Psi_core}}). 
The particular distance along a jet for each panel in Figures~\ref{f:Bp} and \ref{f:Bphi} depends on a selected source properties. For example, for M87 jet, one may set $R_\mathrm{L}=0.08$~pc \citep{Nokhrina19, Kino22}, the viewing angle $17^{\circ}$ and the quasi-parabolic jet shape $r_\mathrm{jet}=0.06z_\mathrm{jet}^{0.57}$ \citep{Nokhrina19}, where $r_\mathrm{jet}$ and $z_\mathrm{jet}$ are in pc. For such a choice of a jet parameters, the panels in Figures~\ref{f:Bp} and \ref{f:Bphi} corresponds to the following distances: $z_\mathrm{proj}=4,\;9,\;41,\;828$~mas. The first two cuts lie inside a M87 parabolic domain, the third cut --- close to the jet shape break and the fourth cut --- in the conical region \citep{Nakamura+18, Nokhrina19}.

The analytical model deviates from the MHD solution only for cross-cuts with $r_\mathrm{jet}\approx R_\mathrm{L}\sqrt{\Gamma_\mathrm{in}\sigma_\mathrm{M}}$, i.e. where the jet structure is central core-dominated (upper left panels in Figures~\ref{f:Bp} and \ref{f:Bphi}). However, this difference is present on the radial scales of $r\lesssim R_\mathrm{L}$ ($0.3$~mas for M87 parameters). The scale of the order of a light cylinder radius constitutes only a minuscule part of a jet width, usually observed, and we do not expect this deviation to affect the results for total intensity or polarization modelling. Outside the light cylinder radius, analytical functions approximate the MHD solution accurately (see upper left panels in Figures~\ref{f:Bp} and \ref{f:Bphi}). 
As the jet propagates and widens, the dense core structure given by Equation~(\ref{Bp2}) is formed. For $r_\mathrm{jet}>R_\mathrm{L}\sqrt{\Gamma_\mathrm{in}\sigma_\mathrm{M}}$ analytical formulae approximate the MHD solution with high accuracy. This is due to our analytical solution aiming to reproduce the jet with a dense core. Outside the light cylinder in the region with $r_\mathrm{jet}\lesssim R_\mathrm{L}\sqrt{\Gamma_\mathrm{in}\sigma_\mathrm{M}}$ and everywhere across the jet for $r_\mathrm{jet}> R_\mathrm{L}\sqrt{\Gamma_\mathrm{in}\sigma_\mathrm{M}}$, the difference between MHD model and analytical approximation becomes $\lesssim 10\%$ for magnetic field amplitudes, $\lesssim 20\%$ for a jet pressure and $\lesssim 10\%$ for total plasma number density. 

Analytical approximation must be used with caution for jet widths $r_\mathrm{jet}\approx R_\mathrm{L}\sqrt{\Gamma_\mathrm{in}\sigma_\mathrm{M}}$ and smaller. This is because the cylindrical approximation may not be applicable for flows not much wider than the light cylinder radius due to importance of longitudinal derivative $\mathrm{d}/\mathrm{d}z$. We plan to address the possible match of analytical approximation with the exact analytical solution in the force-free limit in the following paper.

\section{Plasma bulk motion velocity}
\label{s:vel}

Below we use for brevity the dimensionless jet radius $x=r/R_\mathrm{L}$. As it was shown by \citet{Vlahakis04, Kom09, TMN09}, MHD plasma velocity is equal to the drift velocity in the crossed electric and magnetic fields. Near the axis, where the plasma kinetic energy flux always dominates the Poynting flux, the latter does not hold, and plasma moves with its injection Lorentz factor $\Gamma_\mathrm{in}$. Thus, one obtains the poloidal bulk motion velocity 
\begin{equation}
\frac{v_\mathrm{p}}{c}=
\frac{(1+\varepsilon)x^2}{1+(1+\varepsilon)^2x^2}\approx\frac{x^2}{1+(1+2\varepsilon)x^2},
\label{vp}
\end{equation}
the toroidal velocity 
\begin{equation}
\frac{v_{\varphi}}{c}=\frac{x}{1+(1+\varepsilon)^2x^2}\approx\frac{x}{1+(1+2\varepsilon)x^2},
\label{vphi}
\end{equation}
and the Lorentz factor
\begin{equation}
\Gamma=
\sqrt{\frac{1+x^2(1+\varepsilon)^2}{1+x^2(2\varepsilon+\varepsilon^2)}}\approx\sqrt{\frac{1+x^2}{1+2\varepsilon x^2}}.
\label{G}
\end{equation}

\begin{figure*}
    \centering
    \includegraphics[width=0.45\linewidth]{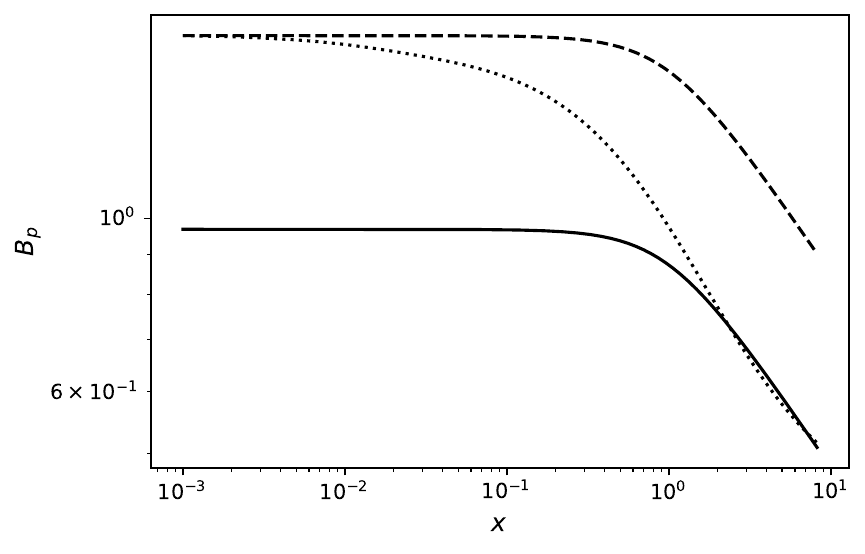} \hspace{0.1cm}
    \includegraphics[width=0.45\linewidth]{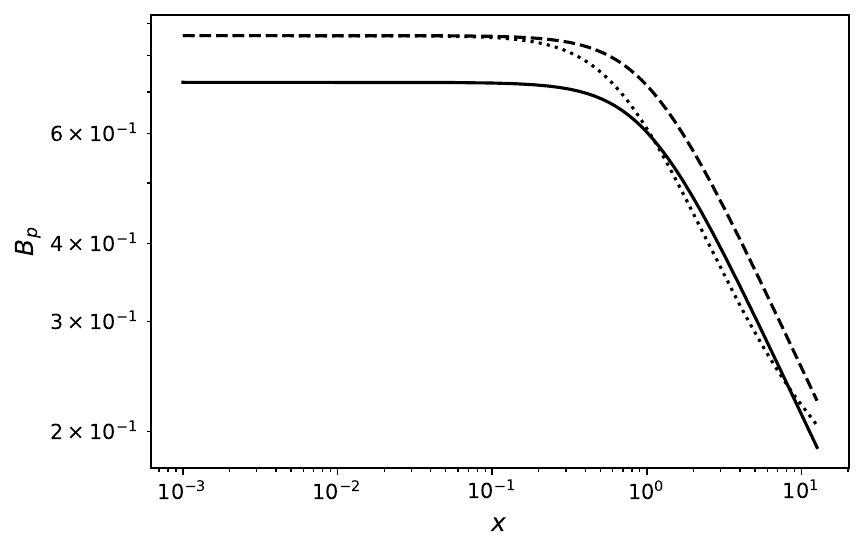} \\ \ \\   
     \includegraphics[width=0.45\linewidth]{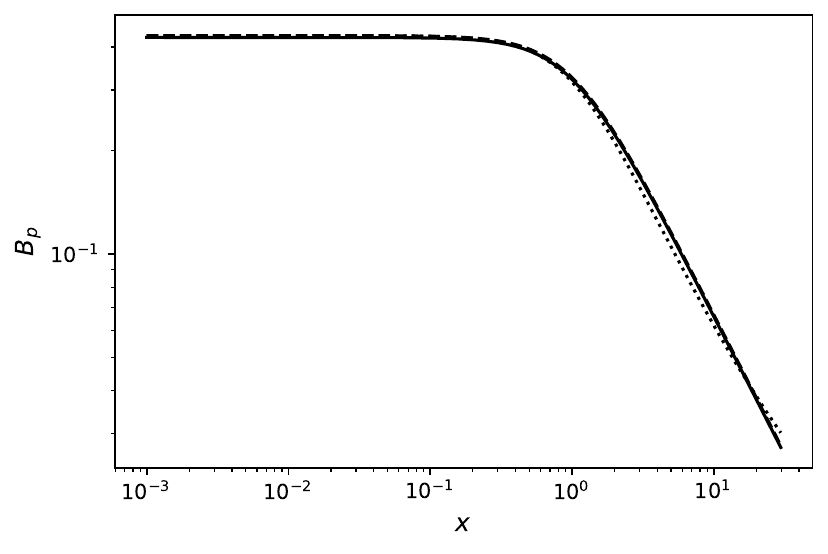} \hspace{0.1cm}
    \includegraphics[width=0.45\linewidth]{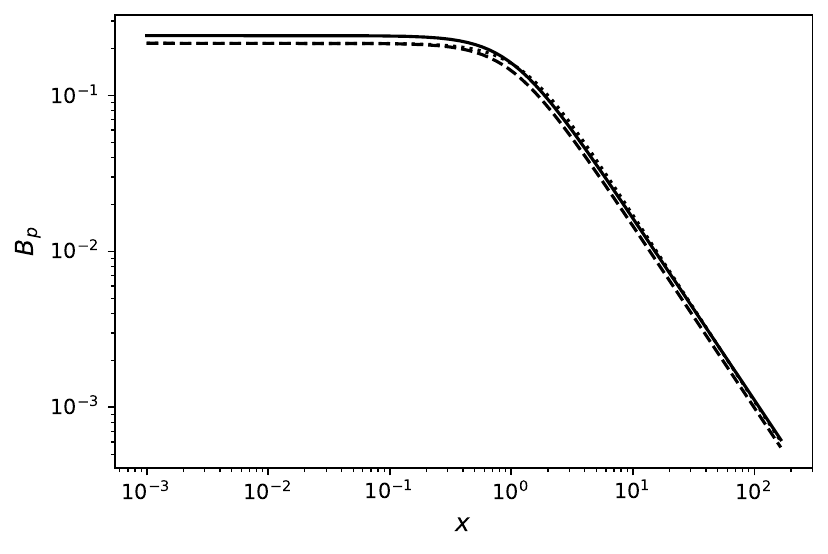}
    \caption{Poloidal magnetic field in units of $\displaystyle\frac{B_\mathrm{L}}{2\sigma_\mathrm{M}}$ across a jet for four different cross-cuts defined by the local jet radius in units of a light cylinder radius (upper left panel corresponds to $x_\mathrm{jet}=8.2$, upper right panel --- $12.6$, bottom left --- $29.7$ and bottom right --- $165.0$)}. Dotted lines --- numerical calculation. Dashed lines --- fits (\ref{Bp2}) and (\ref{alpha1}) with the magnetic field at the axis $B_0$ given by numerical solution. Solid lines --- fits (\ref{Bp2}) and (\ref{alpha1}) with $B_0$ calculated by the conservation of the total magnetic flux (\ref{Psi_core}). Here presented a model with $\sigma_{\rm M}=20$.
    \label{f:Bp}
\end{figure*}

\begin{figure*}
    \centering
    \includegraphics[width=0.45\linewidth]{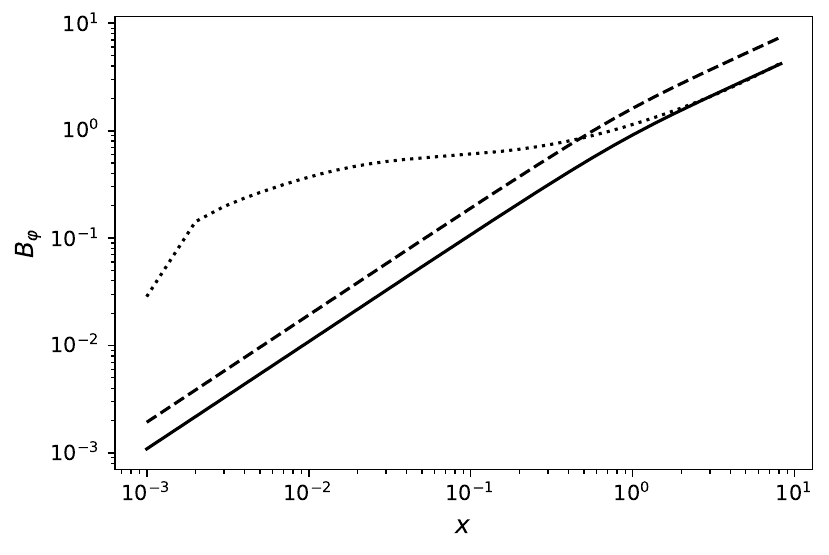} \hspace{0.1cm}
    \includegraphics[width=0.45\linewidth]{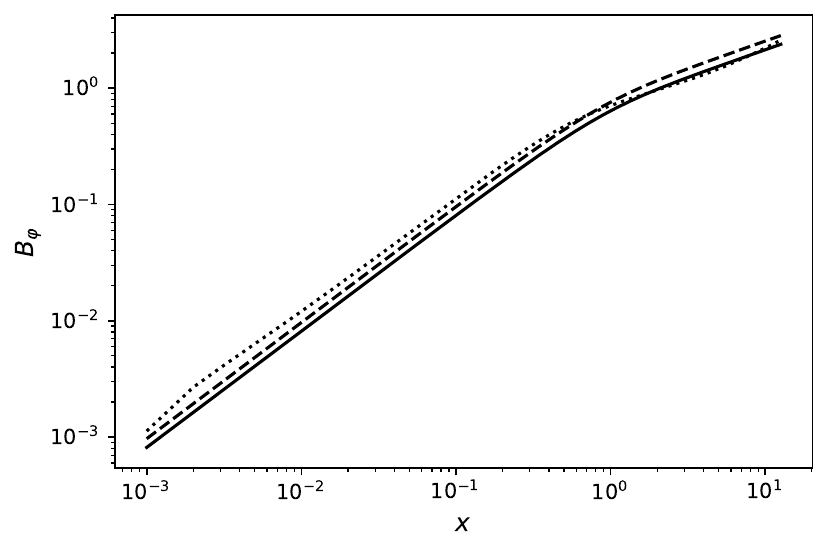} \\ \ \\   
     \includegraphics[width=0.45\linewidth]{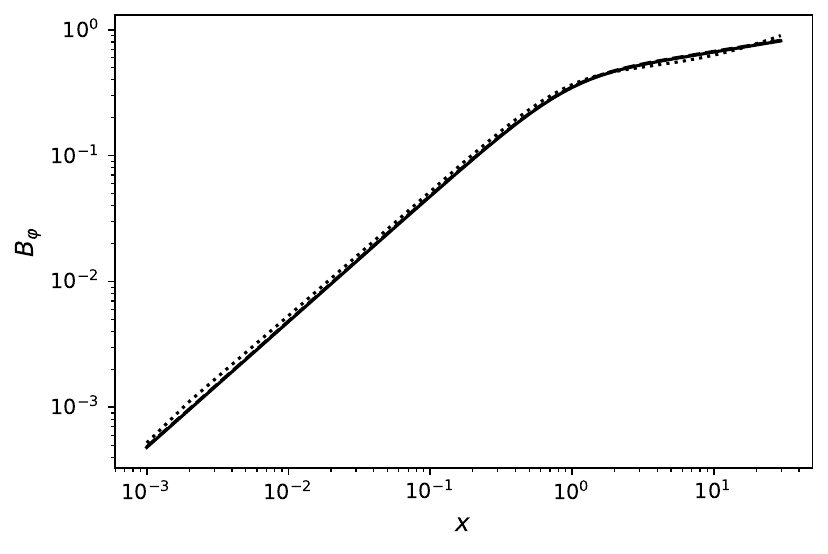} \hspace{0.1cm}
    \includegraphics[width=0.45\linewidth]{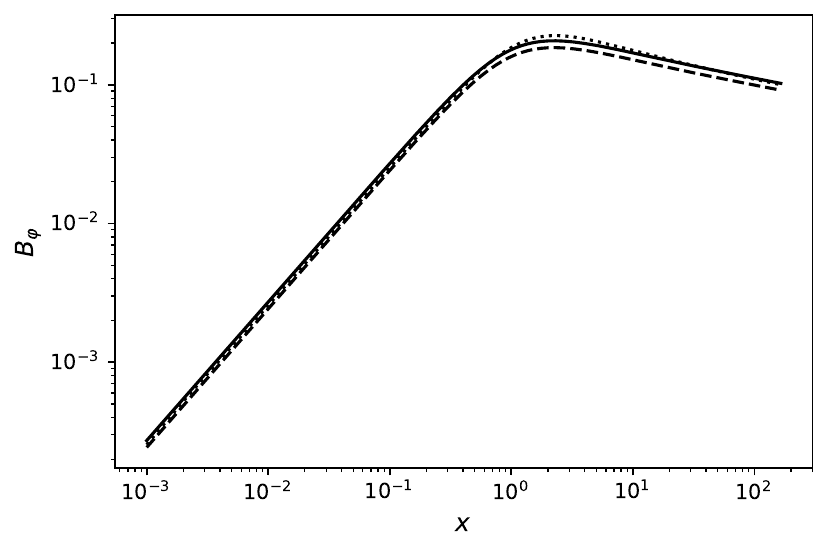}
    \caption{The same as in Figure~\ref{f:Bp} for a toroidal magnetic field component.
    }
    \label{f:Bphi}
\end{figure*}

\begin{figure}
    \centering
    \includegraphics[width=1.0\linewidth]{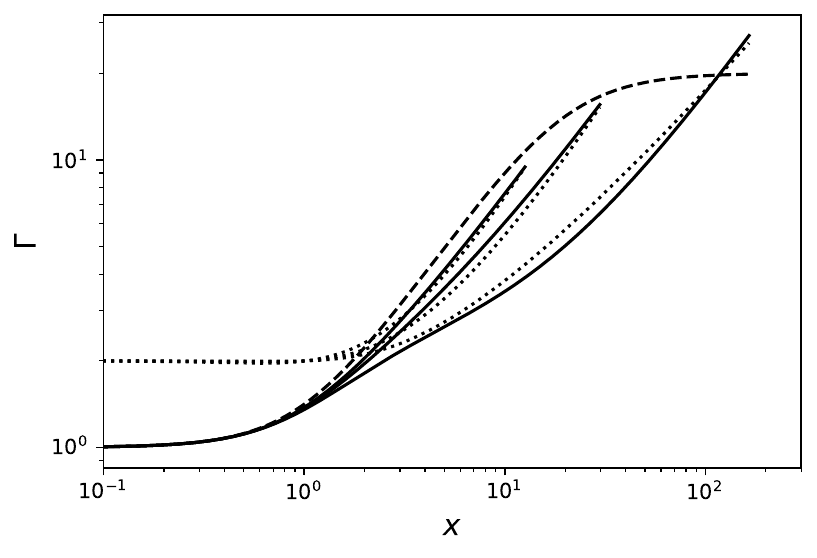} 
    \caption{Bulk plasma motion Lorentz factor across a jet for three different jet cross-cuts. Dotted lines --- numerical calculation. Dashed line --- ideal acceleration given by Equation~(\ref{G}) with a constant $\varepsilon=1/2\sigma_\mathrm{M}^2$. 
    Solid lines --- analytical ansatz given by Equation~(\ref{G}) with a non-constant parameter $\varepsilon$ defined by Equation~(\ref{eps}). Abscissa for each curve designates the jet radius in units of a light cylinder radius.
    }
    \label{f:G}
\end{figure}

\begin{figure*}
    \centering
    \includegraphics[width=0.45\linewidth]{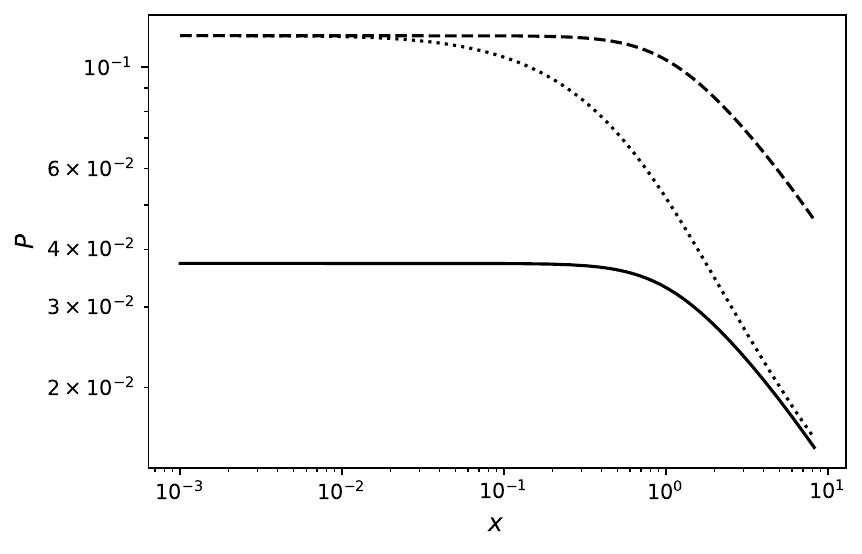} \hspace{0.1cm}
    \includegraphics[width=0.45\linewidth]{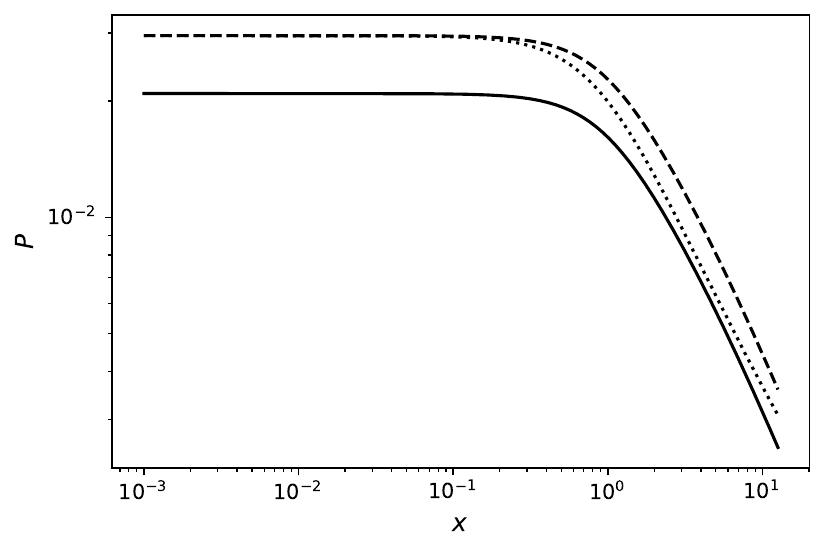} \\ \ \\   
     \includegraphics[width=0.45\linewidth]{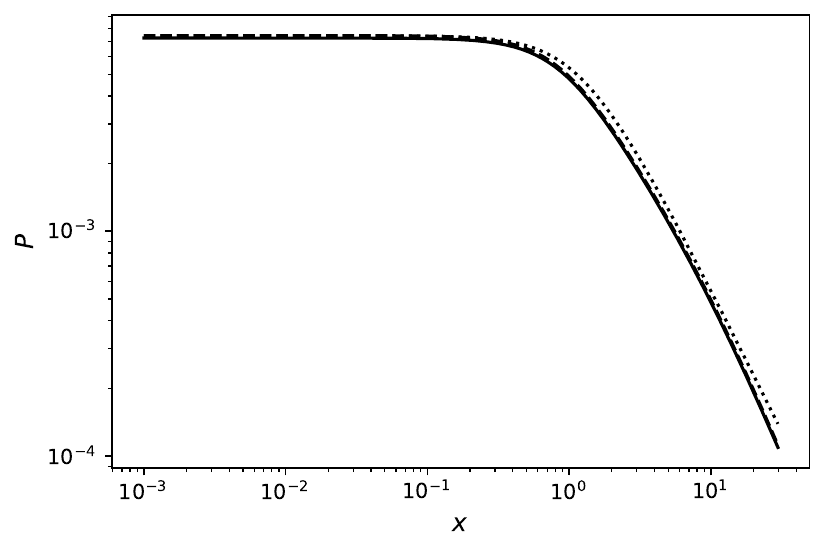} \hspace{0.1cm}
    \includegraphics[width=0.45\linewidth]{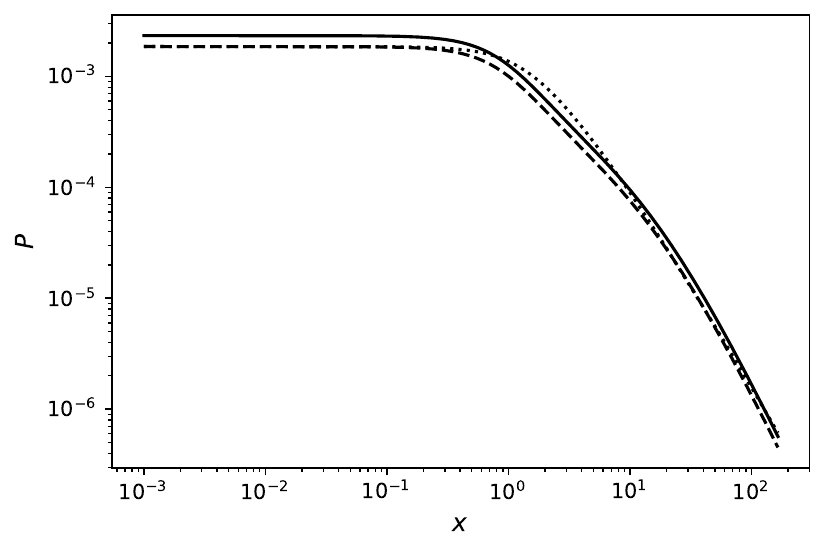}
    \caption{The same as in Figure~\ref{f:Bp} for a total pressure $\displaystyle\frac{B_*^2}{8\pi}$ in units of $\displaystyle\frac{B_\mathrm{L}^2}{4\sigma_\mathrm{M}^2}$.
    }
    \label{f:P}
\end{figure*}

For the ideal linear acceleration
the bulk plasma velocity reaches the saturation at $\Gamma=\Gamma_\mathrm{max}/2=\sigma_\mathrm{M}$ that defines the parameter $\varepsilon$ (B24):
\begin{equation}
\frac{\Gamma_\mathrm{max}}{2}=\frac{1}{\sqrt{2\varepsilon}},
\end{equation}
\begin{equation}
\varepsilon=\frac{1}{2\sigma_\mathrm{M}^2}.
\label{eps_const}
\end{equation}
The expression above works well in the asymptotic regimes. However, in this work we aim to construct an analytical framework for a convenient numerical solution of radiation transfer. As the plasma bulk motion Lorentz factor $\Gamma$ affects Doppler boosting, proper frame magnetic field, plasma number density and jet pressure, we modify the parameter $\varepsilon$ to reproduce as close as possible the behaviour of $\Gamma$ obtained by numerically solving the trans-field Equations~(58)--(59) \citep{Beskin06}.  
We fitted the parameter $\varepsilon$, that defines Lorentz factor $\Gamma$, with the following analytical function:
\begin{equation}
\varepsilon(x,\;x_\mathrm{jet})=\frac{1}{\displaystyle 2\left(\Gamma_\mathrm{in}+\sigma_\mathrm{M}\frac{x}{k(x_\mathrm{jet})}\right)^2},
\label{eps}
\end{equation}
\begin{equation}
k(x_\mathrm{jet})=4.5\sigma_\mathrm{M}\left(\frac{x_\mathrm{jet}}{6\sigma_\mathrm{M}}\right)^{0.65}\left[1+\left(\frac{x_\mathrm{jet}}{6\sigma_\mathrm{M}}\right)^2\right]^{0.15}.
\label{k_eps}
\end{equation}
The dependence of a bulk motion Lorentz factor on a radial distance is presented in Figure~\ref{f:G}. We see that ideal acceleration (dashed curve) approximates satisfactorily only parts of a jet close to its base (with small jet radii). For the farther cross-cuts with larger $r_\mathrm{jet}$ ideal acceleration deviates significantly from the numerical result (dotted curve) in contrast with the analytical approximation given by Equations~(\ref{G}), (\ref{eps}) and (\ref{k_eps}) (solid lines).

Modifying $\varepsilon$ affects the plasma proper frame magnetic field
\begin{equation}
B_{*}^2=B^2-E^2=B_\mathrm{p}^2(1+2\varepsilon x^2).
\label{B_star}
\end{equation}
The analytical approximation allows us to reproduce the MHD solution almost exactly using the modified parameter (\ref{eps}) and (\ref{k_eps}): see Figure~\ref{f:P}, where we show a plasma proper frame magnetic field energy volume density (pressure).

The difference between analytical model and MHD solution for a bulk motion Lorentz factor $\Gamma$ is present for jet parts with $r_\mathrm{jet}\gg R_\mathrm{L}\sigma_\mathrm{M}/2$, i.e. where the jet is far in acceleration saturation regime. This affects the accuracy of other fits, and can be observed between the left and right bottom panels in Figures~\ref{f:Bp}, \ref{f:Bphi}, \ref{f:P} and \ref{f:n_tot}. However, the difference increases very slowly: for example, for M87 parameters the analytical equations approximate the numerical MHD solution with the described above accuracy up to the scales of a few kpc deprojected distance.

For $x\lesssim 1$ numerical calculation provides $\Gamma=\Gamma_\mathrm{in}$ always larger than unity, as it is discussed by \citet{Beskin_KCh_23}. However, in our approach $\Gamma\rightarrow 1$ as $x\rightarrow 0$. This difference provides a minor impact on the results of radiative transfer, because the intensity from the inner jet cylinder with a radius $\sim R_\mathrm{L}$ only is affected within a factor of two. 

\section{Plasma number density}
\label{s:n}

\subsection{Total plasma number density}
\label{ss:n_tot}

In MHD the following relation holds:
\begin{equation}
n\Gamma\frac{v_\mathrm{p}}{c}=\eta B_\mathrm{p}.
\label{n0}
\end{equation}
Here $n$ is a total plasma number density in the plasma proper
frame. Mass-to-magnetic flux function $\eta$ depends on the magnetic flux $\Psi$ only, and is usually taken to be constant. 
The value of this constant is defined as (see Equation~(\ref{sigmaM}))
\begin{equation}
\eta=\frac{\Omega_0^2\Psi_0}{8\pi^2 m_\mathrm{e}c^4\sigma_\mathrm{M}}.
\label{eta}
\end{equation}

This expression must be modified on the scales of a light cylinder radius. The constant $\eta$ corresponds to a non-zero plasma bulk velocity at the jet axis \citep{Beskin06, Lyu09}. However, natural introduction of parameter $\varepsilon$ provides $v_\mathrm{p}(0)=0$ in contradiction with a MHD model. In the Equation~(\ref{n0}) it leads to a total plasma number density divergency as $r\rightarrow 0$. On the other hand, analytical and numerical modelling by \citep{Beskin09, Kom09} demonstrates that $n$ remains constant inside the light cylinder, assuming the value
\begin{equation}
n_\mathrm{core}=\frac{\eta B_0}{\sqrt{\Gamma_\mathrm{in}^2-1}}.
\end{equation}

\begin{figure*}
    \centering
    \includegraphics[width=0.45\linewidth]{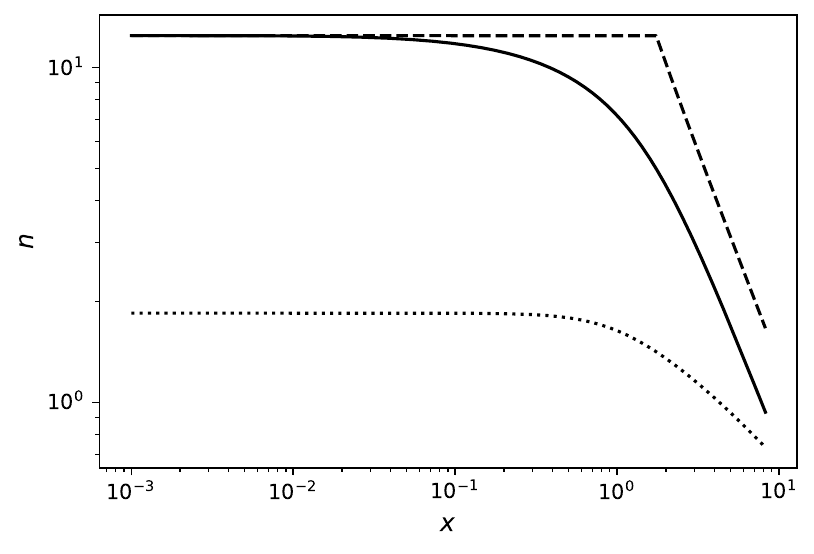} \hspace{0.1cm}
    \includegraphics[width=0.45\linewidth]{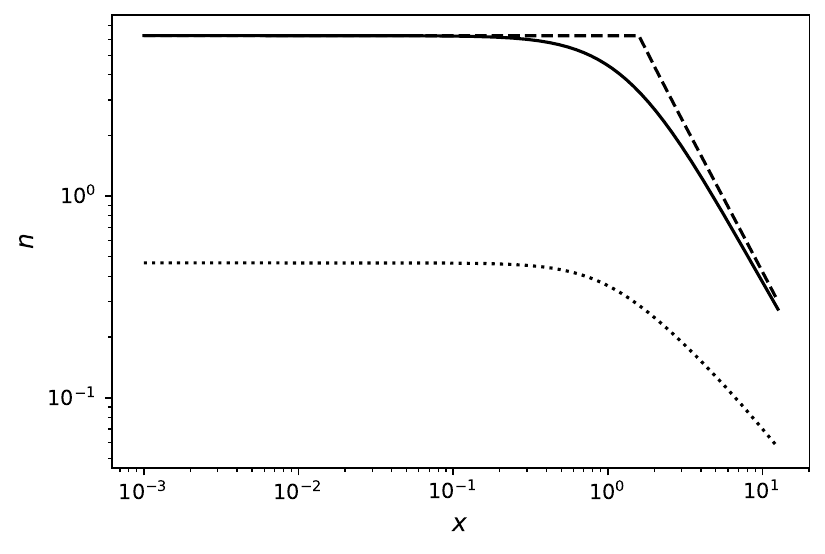} \\ \ \\   
     \includegraphics[width=0.45\linewidth]{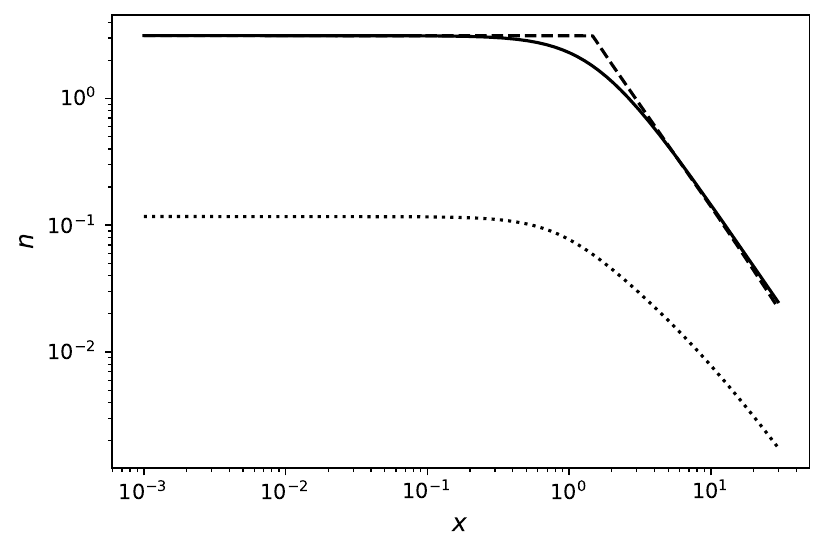} \hspace{0.1cm}
    \includegraphics[width=0.45\linewidth]{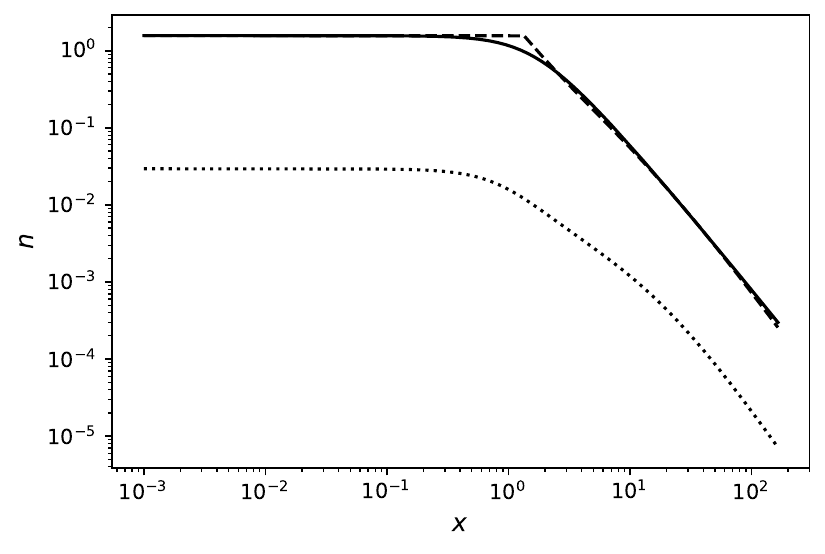}
    \caption{Total MHD plasma number density in units of $m_\mathrm{e}c^2\eta^2$:
    solid lines --- numerical solution; dashed lines --- analytical approximation by Equation~(\ref{n}). Dotted lines: equipartition number density calculated using Equation~(\ref{n_eq}) for $\Sigma=1$. Each panel corresponds to a certain $z$ with a corresponding jet radius from $\sim 9R_\mathrm{L}$ to $\sim 200R_\mathrm{L}$.
    }
    \label{f:n_tot}
\end{figure*}

Thus, we modify the expression~(\ref{n0}) near the axis as follows:
\begin{equation}
n=\mathrm{min}\left\{\frac{\eta B_0}{\sqrt{\Gamma_\mathrm{in}^2-1}};\;\frac{\eta B_\mathrm{p}(x)}{\Gamma(x)v_\mathrm{p}(x)/c}\right\},
\label{n}
\end{equation}
where we use Equations (\ref{Bp1}) or (\ref{Bp2}), (\ref{vp}), (\ref{G}) and (\ref{eta}) to set magnetic field and plasma bulk velocity. The comparison of a total plasma number density, given by Equation~(\ref{n}), and the results of MHD modelling is presented in Figure~\ref{f:n_tot} by dashed and solid lines correspondingly.

Correspondingly, the Alfv\'enic Mach number is defined as
\begin{equation}
{\cal{M}}^2=\frac{4\pi m_\mathrm{e}c^2\eta^2}{n}
\label{M2}
\end{equation}
using the Equations~(\ref{Bp2}), (\ref{vp}), (\ref{G}) and (\ref{n}) reproduces very well the result of numerical solution of MHD equations.

The total plasma number density given by Equation~(\ref{n0}) is connected to the Goldreich--Julian \citep{GJ} number density $n_\mathrm{GJ}$ (in the core frame)
\begin{equation}
n_\mathrm{GJ}=\frac{|\mathbf{\Omega}\cdot \mathbf{B}_\mathrm{p}|}{2\pi c e},
\label{nGJ}
\end{equation}
where $e$ is an absolute value of an electron charge.
The number density in a plasma proper frame is given by \citep{Beskin10}
\begin{equation}
n_\mathrm{plasma}=\lambda\frac{n_\mathrm{GJ}}{\Gamma}.
\label{npl}
\end{equation}
Using the universal relation between the multiplicity parameter $\lambda$, Michel's magnetization parameter and a total jet power $P_\mathrm{j}\approx B_\mathrm{L}^2 R_\mathrm{L}^2 c$ \citep{Beskin10} 
\begin{equation}
\lambda\approx\frac{1}{\sigma_\mathrm{M}}\sqrt{\frac{P_\mathrm{j}e^2}{m_\mathrm{e}^2c^5}},
\end{equation}
one can rewrite Equation~(\ref{npl}) as
\begin{equation}
n_\mathrm{plasma}\approx \frac{v_\mathrm{p}}{c}n.
\end{equation}

\subsection{Equipartition plasma number density}

Suppose that emitting plasma number density proportional to the equipartition number density. In this case $n_\mathrm{em}$ in the plasma proper is defined as
\begin{equation}
n_\mathrm{em}=\frac{B_*^2}{8\pi mc^2\gamma_\mathrm{min}\Lambda\Sigma}.
\label{n_eq}
\end{equation}
Here we assume the power-law energy distribution
\begin{equation}
n_\mathrm{em}(\gamma)=K_\mathrm{e*}\gamma^{-p},\quad \gamma_\mathrm{min}\le \gamma\le\gamma_\mathrm{max}.
\label{ngamma}
\end{equation}
For $p=2$, the emitting plasma number density has the following relation to the amplitude $K_\mathrm{e*}$:
$n_\mathrm{em}=K_\mathrm{e*}/\gamma_\mathrm{min}$. The factor 
$\Lambda=\ln(\gamma_\mathrm{max}/\gamma_\mathrm{min})\approx 10$. The parameter $\Sigma=1$ for equipartition, $\Sigma>1$ for magnetic field dominance and $\Sigma<1$ for emitting plasma energy density dominance. Using the proper frame magnetic field given by Equation~(\ref{B_star}),  
we obtain the equipartition number density
\begin{equation}
n_\mathrm{eq}=\left.n_\mathrm{em}\right|_{\Sigma=1}=\frac{B_\mathrm{p}^2}{8\pi m_\mathrm{e}c^2\gamma_\mathrm{min}\Lambda}\left[1+2\varepsilon(x,\,x_\mathrm{jet})x^2\right].
\label{n_eq}
\end{equation}
Equation~(\ref{n_eq}) means that the equipartition number density (and the proper frame magnetic field) is constant only in the central core-dominated flow (when the jet radius is of the order of $R_\mathrm{L}\sqrt{\Gamma_\mathrm{in}\sigma_\mathrm{M}}$). In the flow with a central core it always decreases towards the jet boundary for the realistic dependence $\alpha(z)$. 

The ratio of an emitting plasma number density to a total number density is equal to
\begin{equation}
\frac{n_\mathrm{em}}{n}=\left(1+2\varepsilon(x,\,x_\mathrm{jet})x^2\right)\frac{B_\mathrm{p}}{B_\mathrm{L}}\Gamma(x)\frac{v_\mathrm{p}}{c}\frac{\sigma_\mathrm{M}}{\gamma_\mathrm{min}\Lambda\Sigma}.
\label{n_ratio}
\end{equation}
The fraction of equipartition plasma number density of the total MHD number density is from a few to tens percent, and its falling with a distance from a jet base. This means that at smaller scales the total plasma number density may not provide for the equipartition number density, leading to overestimating the intensity using the latter. The obtained fraction is exactly the value, needed to explain the observed total jet intensity, as was found by \citet{Frolova23}. 
The behaviour of a plasma number density is shown in Figure~\ref{f:n_tot}. We note that this result holds if the equipartition of the emitting particles is established with the large-scale magnetic field. It could be that only the small-scale (random) component is governing the particles heating, e.g. through turbulence and magnetic reconnection, or some plasma is entrained in the jet volume. However, in this paper we are focused on the MHD jet model without considering such effects.

Equation~(\ref{n_eq}) may be used to assess the emitting plasma number density with $\Sigma\ne 1$ for modelling the brightness temperatures for jets with either magnetic field or emitting plasma number density dominance \citep{Nok17A}. Our approach allows also estimating the emitting plasma number density for the other relations between $n_\mathrm{em}$ and $B_*$, for example, as in \citet{N24, N24b}.

Setting the particular behaviour of an emitting plasma number density allows constructing the spectral flux maps. Some examples of using our analytical approach are presented in Section~\ref{s:intensity}. 

\section{Jet pressure and boundary shape}
\label{s:pressure}

Consider the ambient medium pressure as a power law with respect to the distance from a central engine. Along a jet it is given by
\begin{equation}
P_\mathrm{ext}=P_0\left(\frac{z_0}{z}\right)^{q}
\end{equation}
with $q\approx 2$ for a classical Bondi accretion \citep{QN00, NF11, Park2019}.
Let us assume an equilibrium of jet and ambient medium pressure.

In the central core-dominated flow $B_0>B_\mathrm{cr}$ function $\varepsilon$ can set by Equation~(\ref{eps_const}).
This means that $2\varepsilon x_\mathrm{jet}^2\le \Gamma_\mathrm{in}/\sigma_\mathrm{M}\ll 1$. The jet pressure in this case is defined by the expression
\begin{equation}
P(r_\mathrm{jet})=\frac{\Psi_0^2}{8\pi^3 r_\mathrm{jet}^4},
\label{Pjet1}
\end{equation}
and we obtain the jet boundary shape
\begin{equation}
r_\mathrm{jet}=\left(\frac{\Psi_0^2}{8\pi^3 P_0 z_0^q}\right)^{1/4}z^{q/4}
\end{equation}
close to a parabolic shape for $q\approx 2$. 
Again, this result is applicable only when a jet is collimated well enough on the scales $r_\mathrm{jet}\sim R_\mathrm{L}\sqrt{\Gamma_\mathrm{in}\sigma_\mathrm{M}}$.

In the region, where $B_0<B_\mathrm{cr}$ let us probe the asymptotic behaviour. For $r_\mathrm{jet}\gtrsim R_\mathrm{L}\sqrt{\Gamma_\mathrm{in}\sigma_\mathrm{M}}$
the jet magnetic field pressure $P_\mathrm{jet}\approx B_0^2x_\mathrm{jet}^{-2\alpha}/8\pi$, which leads to the estimate 
\begin{equation}
P_\mathrm{jet}\propto x_\mathrm{jet}^{-4-2\alpha}.
\label{Pjet2}
\end{equation}

At the large distances along a jet, so that $2\varepsilon(x_\mathrm{jet},\;\sigma_\mathrm{M})x_\mathrm{jet}^2\gg\Gamma_\mathrm{in}$ the function $k\approx 0.82\sigma_\mathrm{M}^{0.05}x_\mathrm{jet}^{0.95}$.
In this limit
$P_\mathrm{jet}\approx 2\varepsilon(x_\mathrm{jet},\;\sigma_\mathrm{M})x_\mathrm{jet}^2B^2_0(x_\mathrm{jet})x_\mathrm{jet}^{-2\alpha}$. Substituting in this expression the limiting value for $k$ and definition (\ref{Psi_core}) for $B_0(x_\mathrm{jet})$, we obtain 
\begin{equation}
P_\mathrm{jet}\propto x_\mathrm{jet}^{-2.1},
\label{Pjet3}
\end{equation}
which is consistent with a conical shape for an ambient medium pressure exponent $q\approx 2$.

\begin{figure}
    \centering
    \includegraphics[width=1.0\linewidth]{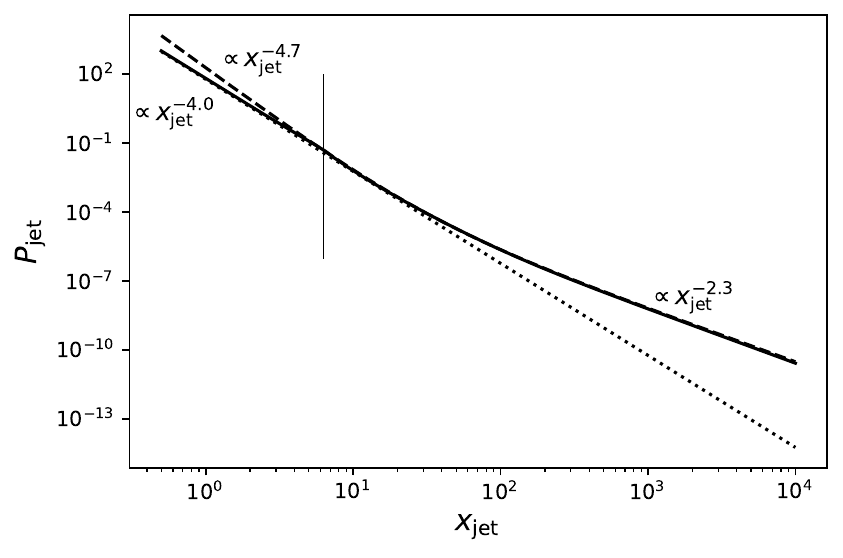} 
    \caption{Dependence of a jet magnetic field at its boundary $P_\mathrm{jet}$ in units of $\displaystyle\frac{B_\mathrm{L}^2}{4\sigma_\mathrm{M}^2}$ on a jet radius in units of a light cylinder radius for $\sigma_\mathrm{M}=20$. Solid line is an analytical modelling. Vertical line corresponds to $r_\mathrm{jet}=R_\mathrm{L}\sqrt{\Gamma_\mathrm{in}\sigma_\mathrm{M}}$. 
    Dotted line represents an asymptotic at the smallest radii $P_\mathrm{jet}\propto r_\mathrm{jet}^{-4}$. Dashed line --- a fit for $r_\mathrm{jet}>R_\mathrm{L}\sqrt{\Gamma_\mathrm{in}\sigma_\mathrm{M}}$ with asymptotics $P_\mathrm{jet}\propto r_\mathrm{jet}^{-4.7}$ and $P_\mathrm{jet}\propto r_\mathrm{jet}^{-2.3}$.
    }
    \label{f:Pjet}
\end{figure}

The behaviour of a jet pressure at different scales is presented in Figure~\ref{f:Pjet}.
For $r_\mathrm{jet}>R_\mathrm{L}\sqrt{\Gamma_\mathrm{in}\sigma_\mathrm{M}}$ the jet pressure is approximated by the broken power-law
\begin{equation}
P_\mathrm{jet}=1.13\times 10^{-3}\frac{B_\mathrm{L}^2}{\sigma_\mathrm{M}^{3.8}}\left(\frac{x_\mathrm{jet}}{1.5\sigma_\mathrm{M}}\right)^{-b}\left[1+\left(\frac{x_\mathrm{jet}}{1.5\sigma_\mathrm{M}}\right)\right]^{b-2.3},
\label{Pjet_fit}
\end{equation}
where $b=5.39\sigma_\mathrm{M}^{-0.046}$ depends weakly on an initial jet magnetization.
Equations~(\ref{Pjet1}) and (\ref{Pjet_fit}) may be used to infer the jet parameters from the observed jet shapes.

\section{Emissivity across a jet}
\label{s:intensity}

\vskip 5mm

In the optically thin regime spectral intensity $I_{\nu*}$ in a plasma proper frame is proportional to the emission coefficient $j_{\nu*}$ (see, e.g., Equation~(35) in B24): $I_{\nu*}=\int j_{\nu*} ds_*$, where $s_*$ is a length along a line of sight in a plasma proper frame. Let us introduce the Doppler-corrected emissivity $I$ so that it includes both the emission coefficient $j_{\nu*}$ and the factor of transformation into a core frame: $I_{\nu*}=\int Ids$. The emissivity coefficient is proportional to $B_*^{(p+1)/2}$ and to the proper frame emitting plasma number density $n_\mathrm{em}$. For the equipartition assumption $j_{\nu*}\propto B_*^{2+(p+1)/2}$. Due to a transition into a core frame, the value $I$ obtains a factor $\delta^{2+(p-1)/2}$ \citep[see e.g.][]{Lyutikov2005}. For a jet with negligible viewing angle, Doppler factor $\delta=2\Gamma\propto x$. Finally, the Doppler-corrected emissivity $I\propto \delta^{2+(p-1)/2}B_*^{2+(p+1)/2}$.

\begin{figure*}
    \centering
    \includegraphics[width=0.45\linewidth]{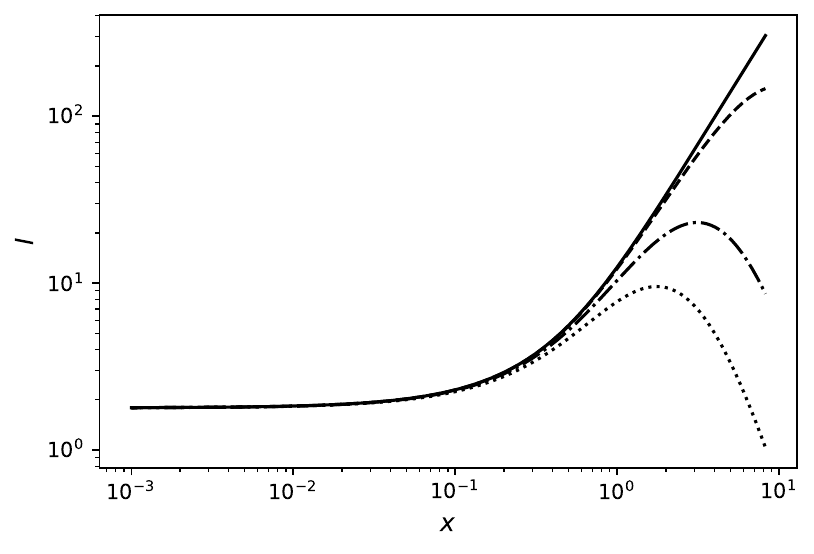} \hspace{0.1cm}
    \includegraphics[width=0.45\linewidth]{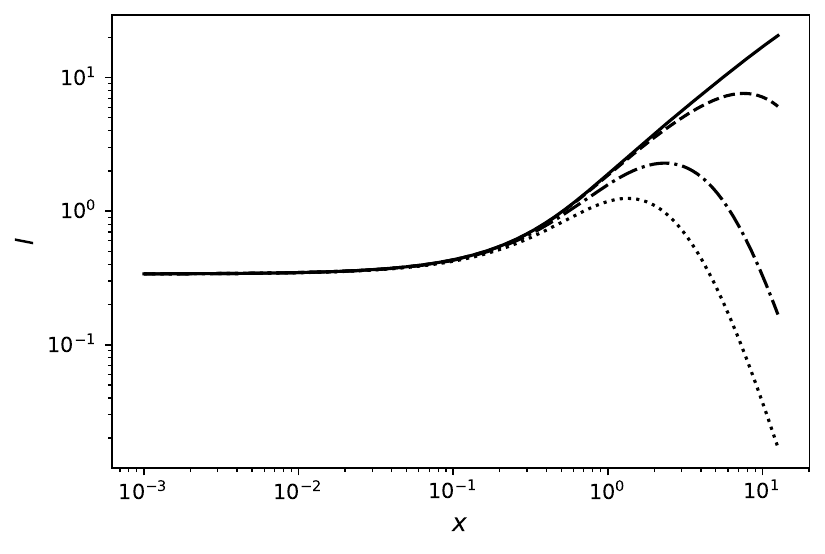} \\ \ \\   
     \includegraphics[width=0.45\linewidth]{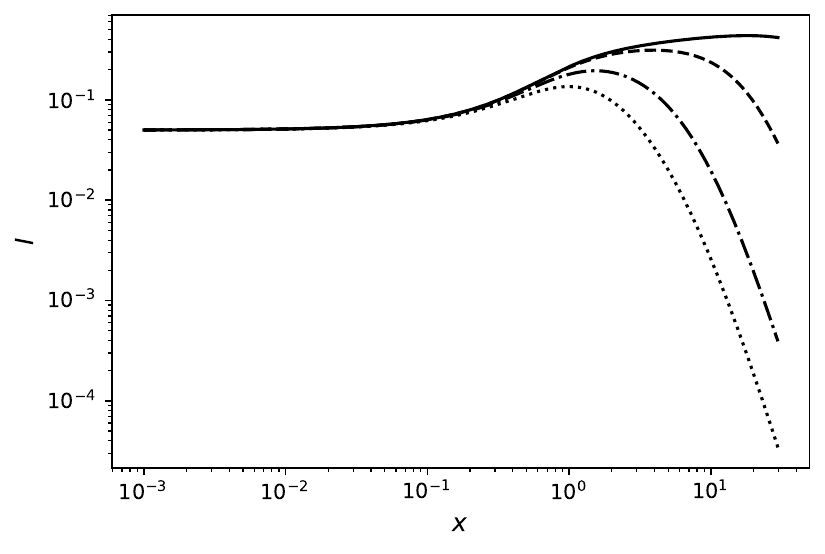} \hspace{0.1cm}
    \includegraphics[width=0.45\linewidth]{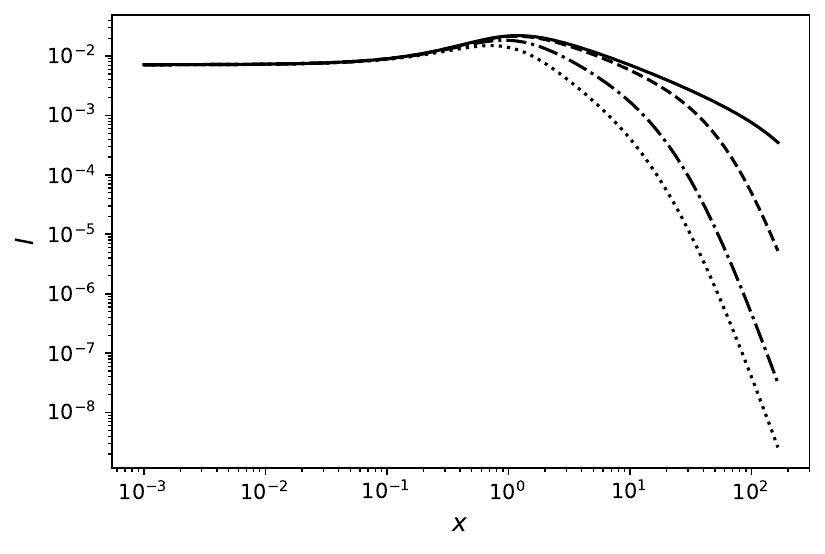}
    \caption{Dimensionless Doppler-corrected emissivity $I$ for the same jet cross-cuts as in Figures~\ref{f:Bp}--\ref{f:n_tot}. The chosen parameters are $p=2$, $\sigma_\mathrm{M}=20$. Solid, dashed, dashed-dotted and dotted lines correspond to the viewing angles $1^{\circ}$, $5^{\circ}$, $15^{\circ}$, $25^{\circ}$ respectively.
    }
    \label{f:I}
\end{figure*}

Employing this assumption for a Doppler factor, let us qualitatively estimate the behaviour of the Doppler-corrected emissivity $I$ for an `ideal' linear acceleration with saturation described by $\varepsilon=1/2\sigma_\mathrm{M}^2$. 
Using Equations~(\ref{G}) and (\ref{B_star}), we obtain the following asymptotic behaviour:
\begin{equation}
I 
\propto
\left\{
\begin{array}{l}
\displaystyle \mathrm{const},\quad x<\sqrt{\Gamma_\mathrm{in}\sigma_\mathrm{M}},\\ \ \\
\displaystyle x^{\left[(3+p)-\alpha(5+p)\right]/2}, \quad 1\ll x\ll\sigma_\mathrm{M},\\ \ \\
\displaystyle x^{(1-\alpha)(5+p)/2}, \quad x\gg\sigma_\mathrm{M}.
\end{array}
\right.
\label{S}
\end{equation}
The exponent $(1-\alpha)(5+p)/2$ in the last line is always negative, as we see from Figure~\ref{f:alpha}. So, $I$ from the outer parts of a jet decreases. The exponent $\left[(3+p)-\alpha(5+p)\right]/2$ is positive for $\alpha<(3+p)/(5+p)$, which is always the case for $x<\sigma_\mathrm{M}$. Thus, for these jet parts we may expect that the Doppler-corrected emissivity $I$ increases towards the radius of the order of a few $R_\mathrm{L}\sigma_\mathrm{M}$. This means a possible appearance of a limb brightening in the sub-parsec parts of a jet found in B24. The Lorentz factor behaviour, consistent with MHD model with $\varepsilon=\varepsilon(x,\;x_\mathrm{jet})$ instead of a constant, follows this trend qualitatively.
We plot the function $I$ for different jet cross-cuts and at different viewing angles (see Figure~\ref{f:I}). We see, that, in general, we should expect growth of a Doppler-corrected emissivity as the line of sight moves from the jet axis for an effectively accelerating parts of a jet observed at the small enough viewing angle $\lesssim\Gamma^{-1}$. Otherwise, the limb brightening may be ensured only for a specific spacial distribution of emitting relativistic plasma.

Here we must note that described above picture is valid only for an isotropic distribution of emitting particles over the pitch angles. The evolution of plasma distribution due to an adiabatic invariant conservation \citep{Beskin2023} or appearance of anisotropy in a pitch angle \citep{Tsunetoe2025,Beskin2025} may alter the discussed above spectral emissivity.

\begin{figure*}
    \centering
    \includegraphics[width=0.95\linewidth]{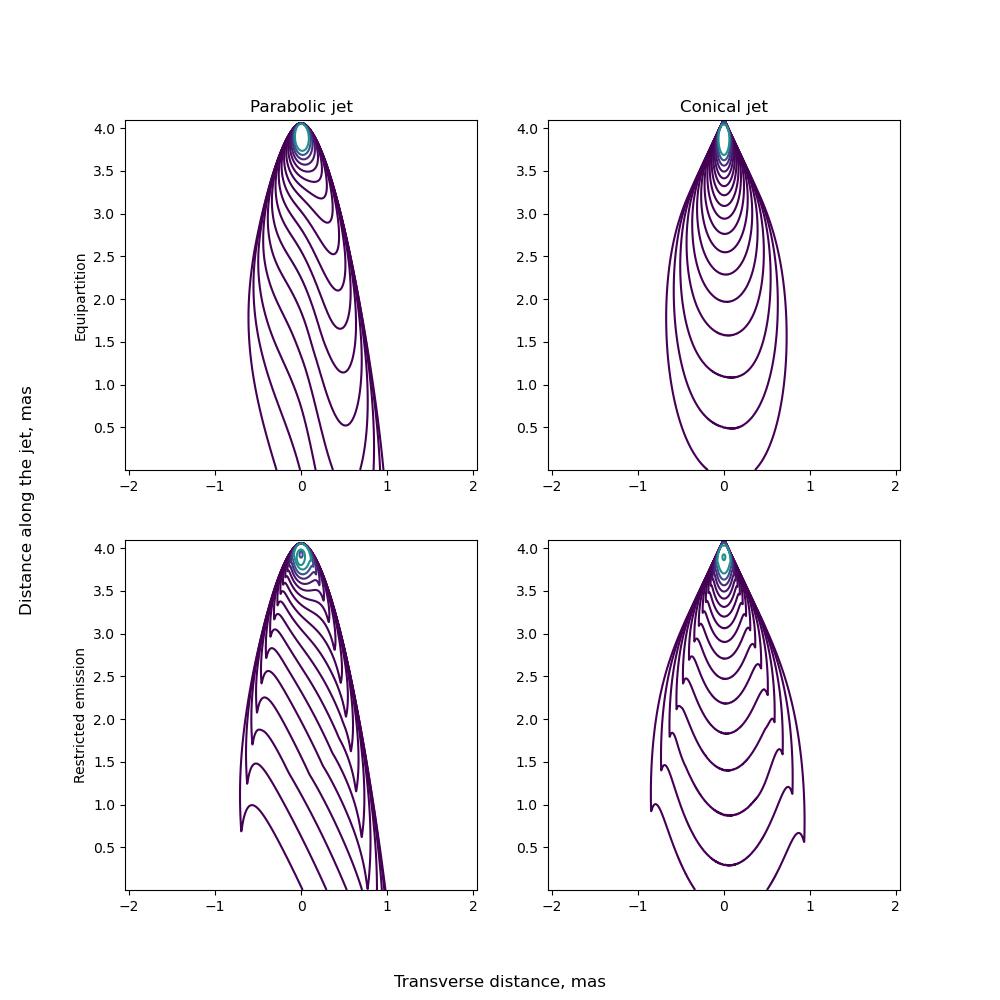}
    \caption{The contours maps for $R_{\rm L} = 0.01$ pc. Left: parabolic geometry with the viewing angle $15^{\circ}$, right: conical geometry with the viewing angle $1^{\circ}$, top: the emitting particles are distributed according to the equipartition assumption, bottom: the particles with $\Psi \geqslant 0.75 \Psi_0$ emit. The pixel is $0.001$ mas, the map is $4096$ by $4096$ pixels, the map total spectral flux is $1$ Jy. The highest contour level is the map maximal value, successive contours are drawn as $c_n = c_{n-1}/2$.}
    \label{fig:RL0.01_contours}
\end{figure*}

\begin{figure*}
    \centering
    \includegraphics[width=0.95\linewidth]{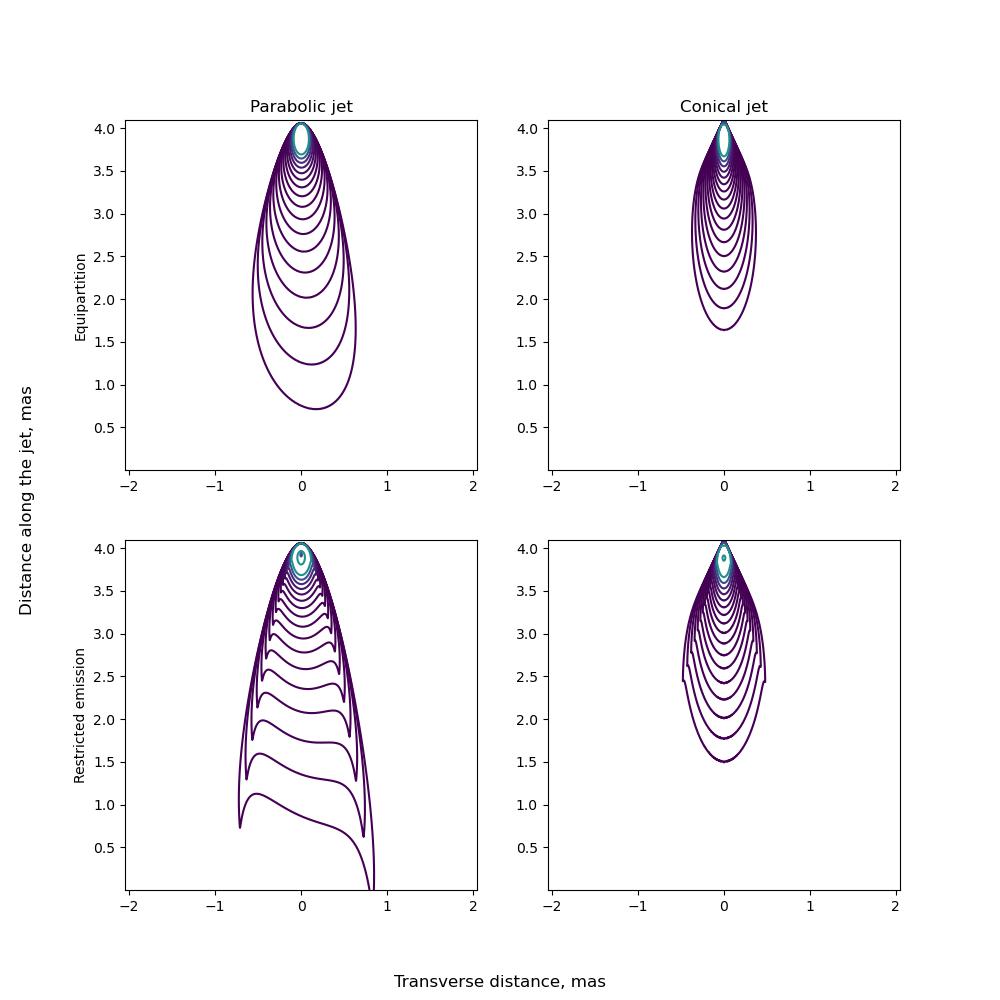}
    \caption{The same as \autoref{fig:RL0.01_contours} but for $R_{\rm L} = 0.1$ pc.}
    \label{fig:RL0.1_contours}
\end{figure*}

The spectral intensity in addition to the emissivity includes the effects of a emitting plasma geometrical depth and optical depth. In order to test fully the model, we conducted several radiation transfer simulations. First, we set the outer geometry of the jet. We considered a M87-like parabolic geometry ($r_\mathrm{jet}=0.07z^{0.57}$, \citet{Nokhrina19}) and a conical geometry with an intrinsic opening angle $1.2^{\circ}$ which is the median value in the MOJAVE survey \citep{MOJAVE_XIV}. We set the observational frequency of $15$ GHz and the viewing angle is $15^{\circ}$ for parabolic geometry and $1^{\circ}$ for conical geometry. We calculated magnetic field components and plasma bulk motion velocity using equations (\ref{Bp0})--(\ref{E0}) and (\ref{Psi_core})--(\ref{vphi}), and we considered two possible spatial emitting particle number density distributions: a. the equipartition between the emitting particles and magnetic field energy, b. the particles with the magnetic flux above the threshold value emit. In this case, the emitting plasma is distributed according to \eqref{n_eq}. As the threshold value, we chose $75$ per cent of the total magnetic flux in the jet, $\Psi_0$. This is the fiducial value to explore a general picture of cut off emitting particles distributions. We assume an isotropic power-law distribution of emitting particles given by the Equation~(\ref{ngamma}) with $\gamma_{\rm min} = 10$ and index $p = 2.5$ which corresponds to a spectral index $-0.75$ \citep{Hovatta2014}. We considered two different values of light cylinder radius, $0.01$ (\autoref{fig:RL0.01_contours}) and $0.1$ pc (\autoref{fig:RL0.1_contours}), to show the impact of black hole spin/mass variation. We set the total magnetic flux $\Psi_0$ so it normalizes the resulting intensity map to the total spectral flux of $1$ Jy. This value conveys the correct order of magnitude for a wide range of sources. All the resulting $\Psi_0$ values have the order of $10^{32}-10^{33}$ G cm$^2$. For uniformity, both for the parabolic and conical geometry we set the pixel of $0.001$ mas and the ratio $0.08$ pc/mas. The images are $4096$ by $4096$ pixels. For each pixel, we solve the radiative transfer equation for $I_{\nu}$ that has the form
\begin{equation}
    \frac{{\rm d}I_{\nu}}{{\rm d}s} = j_{\nu} - \text{\ae}_{\nu} I_{\nu},
\end{equation}
where $I_{\nu}$ is the spectral intensity, $j_{\nu}$ and $\text{\ae}_{\nu}$ are the emission and absorption coefficients \citep{GinzburgSyrovatskii65} and $s$ parametrizes the line of sight. The equation is given in the observer frame of reference while the expressions for $j_{\nu}$ and $\text{\ae}_{\nu}$ are defined in the plasma proper frame. To recalculate the coefficients to the observer frame, we use the Lorentz invariants $I_{\nu}/\nu^3$, $j_{\nu}/\nu^2$ and $\text{\ae}_{\nu}\nu$ \citep{RL79}, and $j_{\nu}$ and $\text{\ae}_{\nu}$ obtain factors $\delta^{(3+p)/2}$ and $\delta^{(2+p)/2}$ correspondingly. 

We plot the maps in contours. For contours plotting, the highest contour level is the map maximal value, successive contours are drawn as $c_n = c_{n-1}/2$. Their number is chosen visually. We see, that, indeed, although the emissivity nearby the jet base grows towards the jet boundaries, the low geometrical depth suppresses this effect, which is somewhat pronounced only in the left upper panel of the Figure~\ref{fig:RL0.01_contours}. Mimicking the growing towards the boundary jet loading with emitting plasma used by \citet{Hir25} (lower panels in Figures~\ref{fig:RL0.01_contours} and \ref{fig:RL0.1_contours}), we obtain both limb brightened and triple ridge brightened patterns. 

\section{Discussion and Conclusions}
\label{s:Conc}

We propose an analytical MHD model that can be conveniently used to calculate numerically the synchrotron self-absorbed emission of relativistic jets. The model is applicable for well collimated flows. The full MHD model can be used far from the jet base, where the force free assumption is no more applicable. It allows using any desirable resolution of a jet structure, including the light cylinder radius that expected to be of the order of 10$-$100 gravitational radii \citep{McKinney06, Nokhrina19, Kino22}, which provides a few of $10^{-3}$~pc for a black hole with mass $10^9\;M_{\odot}$. The presented analytical model provides a fine control of a jet parameters, thus enabling us to study their impact on an observed jet image. It is important that it reproduces accurately not only the electro-magnetic field profiles along and across a jet, but a magnetic field in a plasma proper frame as well. Section~\ref{s:sec1} fully describes the analytical approximations for an electric and a magnetic fields, while Section~\ref{s:vel} --- bulk motion velocity field. The full description of a jet structure allows calculating full intensity and polarization maps.

Our model makes straightforward setting up the analytical jet structure.
Electric and magnetic fields are defined by Equations~(\ref{Bp0})--(\ref{Bphi0}), with the total magnetic flux given by Equations~(\ref{Psi_stream}) or (\ref{Psi_core}). The exponent of a poloidal field is calculated using (\ref{alpha1}) and (\ref{params}), and the magnitude of a poloidal magnetic field at the axis using (\ref{Bp1}) or (\ref{B0_core}).

The velocity field is calculated using Equations~(\ref{vp})--(\ref{G}) with functions~(\ref{eps}) and (\ref{k_eps}). The total and emitting plasma number density are defined by Equation~(\ref{n}) and (\ref{n_eq}).

There can two ways to define a jet boundary: either setting it explicitly as a function $z(r_\mathrm{jet})$, or by defining an ambient medium pressure $P(z)$ and using the curve (\ref{Pjet_fit}).

The analytical approximation presented here allows to apply different models for emitting plasma spatial distribution to create the intensity and polarization maps. In particular, our model provides the expression for a total plasma number density~(\ref{n}), which is very close to the Goldreich--Julian number density~(\ref{nGJ}) for a relativistic flow.  
A model for electromagnetic field and velocity distribution provides a way to calculate the equipartition plasma number density~(\ref{n_eq}). We observe (Figure~\ref{f:n_tot}) that for wider parts of a jet the latter constitutes about a few per cent of a total plasma number density. 
However, closer to the central source the equipartition plasma number density constitutes about a half of a total plasma number density at the jet boundary (see left upper panel in Figure~\ref{f:n_tot}). 
This means that for the innermost jet parts the possible supply of equipartition plasma number density may be restricted by the total plasma availability in an MHD flow. Interesting, that the decrease of the core brightness temperatures close to the jet base \citep{2016ApJ...826..135L} could be associated with this. A detailed model for electric and magnetic field distribution provides also an instrument to calculate an emitting plasma number density proportional to a square of a proper electric current density (see Appendix~\ref{a:j_prime}), proposed to explain an observed characteristic jet polarization pattern \citep{Lyutikov2005, Frolova23}. The emitting plasma number density can be easily calculated also for non-equipartition sate \citep{Nok17A, N24, N24b}. 

A fine tuning of a proper frame magnetic field $B_*$ by analytics allows modelling a jet shape by setting an ambient medium pressure as a function of a distance along a jet $P_\mathrm{ext}(z)$. Indeed, we obtained the parabolic boundary shape not only when a jet is magnetically-dominated (dashed line in Figure~\ref{f:Pjet}), but also for a central core-dominated flow, for which Equation~(\ref{Pjet1}) holds. However, this is an ideal jet boundary defined by a total magnetic flux conservation. There may be deviations in total intensity maps.

We must note here, that the model with a constant angular velocity $\Omega_\mathrm{F}=\mathrm{const}$ and linear functions of an energy $E(\Psi)$ and angular momentum $L(\Psi)$ density (see Equations~(52) and (53) in \citet{Beskin06}) corresponds to a maximum energy extraction via Blandford--Znajek process \citep{BZ-77, ModSik96, Hir25} towards a jet boundary. However, this does not necessarily converts into a limb brightening effect. Total spectral intensity is defined by magnitudes of a magnetic field and emitting plasma number density in a plasma proper frame together with a Doppler--boosting. Magnetic field magnitude $B_*$ decreases towards a jet boundary (see Figure~\ref{f:P}). Indeed, for a magnetically-dominated flow with effective acceleration, expected in jets that reach the observed Lorentz factors \citep{MOJAVE_XVII}, the proper frame magnetic field is of the order of $B_\mathrm{p}$ \citep{Vlahakis04, Kom09}. Thus, $B_*$ decreases always for a core flow, which is formed for reasonable supporting a jet ambient medium pressure greater than $B_\mathrm{cr}^2/8\pi$ \citep{McKinney06, Kom07, Beskin09, Lyu09, Porth11, Tchekhovskoy_11}. In a case of plasma-dominated outflow, a proper frame magnetic field is of the order of $\sqrt{B_{\varphi}^2-E^2}\approx\sqrt{2\varepsilon}B_\mathrm{p}r/R_\mathrm{L}$ for a constant angular velocity $\Omega_\mathrm{F}$ and $B_*\approx\sqrt{2\varepsilon}B_\mathrm{p}\omega r/R_\mathrm{L}$ for $\omega=\Omega_\mathrm{F}(\Psi)/\Omega_0$. In both case, as we expect the decreasing function $\omega(\Psi)$, $B_\mathrm{p}r$ decreases towards the jet boundary for $\alpha>1$ (see Equation~\ref{Bp2} and Figure~\ref{f:alpha}). Thus, if the emitting plasma is distributed corresponding to the equipartition assumption~(\ref{n_eq}), the emissivity in a plasma proper frame drops with a radius. The Doppler boosting can made up for such a decrease, but only for a small enough viewing angles and close to a jet base (see Equation~(\ref{S}) and Figure~\ref{f:I}). Thus, the relativistic jets, described by a constant angular velocity, equipartition emitting plasma number density and observed at small viewing angles, are expected to be limb brightened close to its base and spine-brightened downstream. The coincidence in M87 jet of a detected bulk plasma velocity saturation by \citet{Mertens16} (Figure~16) with the span of an observed limb brightened part of a jet by by \citet{Walker18} (Figure~7) supports our conclusion, although a detailed analysis is needed in this case. Further corroboration is an apparent absence of a limb brightening effect in M87 counter-jet \citep{Walker18}, which is explained by the absence of a Doppler-boosting that enhances a laboratory frame intensity despite the decreasing emissivity in a plasma proper frame.

Another way to account for the limb brightening is assuming a non-constant mass-to-magnetic flux function $\eta$. Here we should make two remarks about our assumption of $\eta=\mathrm{const}$. (i.) Analytical modelling by \citet{Beskin_KCh_23} shows, that even if $v_\mathrm{p}$ and $\eta$ vanish at the jet axis, the central core $B_\mathrm{p}\approx \mathrm{const}$ still forms within a light cylinder. If we assume that all the plasma, with number density given by $n_\mathrm{tot}$, or equipartition plasma number density $n_\mathrm{eq}$ emit, such emission will be pronounced along a jet spine or even dominate across a jet \citep[see, e.g. discussion by][]{Frolova23}. (ii.) Increasing function $\eta(\Psi)$ represents a mass loading towards a jet boundary, so it is important to explore jet models with such an effect. We plan to address this issue in the following work. Here we must note that we mimic the mass loading effect at the jet boundary by assuming that plasma does not emit in the jet central region, and that all the plasma emits around a jet boundary (see Figures~\ref{fig:RL0.01_contours} and \ref{fig:RL0.1_contours}).

\section*{Acknowledgements}

We thank the anonymous referee for suggestions that helped to
improve the paper. E.E.N. and V.A.F. acknowledge financial support by the Foundation for the Advancement of Theoretical
Physics and Mathematics ``BASIS''. This research made use of the data from the MOJAVE database\footnote{\url{http://www.physics.purdue.edu/MOJAVE/}} which is maintained by the MOJAVE team \citep{MOJAVE_XV}.
This research made use of NASA's Astrophysics Data System.

\section*{Data availability}

There is no new data associated with the results presented in the paper. All the previously published data has the proper references.

\bibliographystyle{mnras}
\bibliography{nee1}

\appendix

\section{Electric current density in a plasma proper frame}
\label{a:j_prime}

To calculate the electric current density in a plasma proper frame we use the $4$-current with the components $j^{\mu}=\{c\rho,\,{\textbf{j}}\}$ in a laboratory frame and apply the Lorentz transformations. Here $\rho$ is a charge density, and an electric current and charge densities are defined by the Maxwell's equations.
Below we use the designations 
\begin{equation}
\Psi_r=\frac{\partial\Psi}{\partial r};\quad \Psi_z=\frac{\partial\Psi}{\partial z}.
\end{equation}

The components of $j^{\mu}$ are the following.
\begin{equation}
c\rho=-\frac{c}{8\pi^2R_\mathrm{L}}\left(\frac{\Psi_r}{r}+\Psi_{rr}+\Psi_{zz}\right),
\end{equation}
\begin{equation}
\begin{array}{l}
\displaystyle j_r=-\frac{c}{8\pi^2R_\mathrm{L}\sqrt{\Psi_r^2+\Psi_z^2}}\left(\Psi_r\Psi_{rz}+\Psi_z\Psi_{zz}\right)-\\ \ \\
\displaystyle-\frac{\partial\varepsilon}{\partial z}\frac{c}{8\pi^2 R_\mathrm{L}}\sqrt{\Psi_r^2+\Psi_z^2},
\end{array}
\end{equation}
\begin{equation}
j_\varphi=-\frac{c}{8\pi^2r}\left(\Psi_{rr}+\Psi_{zz}-\frac{\Psi_r}{r}\right),
\end{equation}
\begin{equation}
\begin{array}{l}
\displaystyle j_z=-\frac{c(1+\varepsilon)}{8\pi^2R_\mathrm{L}\sqrt{\Psi_r^2+\Psi_z^2}}\left[\left(\Psi_r^2+\Psi_z^2\right)\left(\frac{1}{r}+\right.\right.\\ \ \\
\displaystyle\left.\left.+\frac{1}{1+\varepsilon}\cdot\frac{d\varepsilon}{dr}\right)+\left(\Psi_r\Psi_{rr}+\Psi_z\Psi_{zr}\right)\right].
\end{array}
\end{equation}
Derivative $\partial\varepsilon/\partial z$ is non-trivial due to the dependence $\varepsilon(x_\mathrm{jet})=\varepsilon(x_\mathrm{jet}(z))$.

The drift velocity components are listed below.
\begin{equation}
\left.\left(\frac{v}{c}\right)\right|_{r}=\beta_r=-\frac{(1+\varepsilon)x^2}{1+(1+\varepsilon)^2x^2}\frac{\Psi_z}{\sqrt{\Psi_r^2+\Psi_z^2}},
\end{equation}
\begin{equation}
\beta_\varphi=\frac{x}{1+(1+\varepsilon)^2x^2},
\end{equation}
\begin{equation}
\beta_z=\frac{(1+\varepsilon)x^2}{1+(1+\varepsilon)^2x^2}\frac{\Psi_r}{\sqrt{\Psi_r^2+\Psi_z^2}}.
\end{equation}
They provide the Equation~(\ref{vp}) with the Lorentz factor described by the Equation~(\ref{G}). 

Introducing the projecting on a velocity vector $\boldsymbol\beta=\{\beta_r,\,\beta_\varphi,\,\beta_z\}$ operator acting on an arbitrary vector $\textbf{A}$ as
\begin{equation}
\widehat{\boldsymbol\beta\boldsymbol\beta}{\textbf{A}}=\frac{\left(\textbf{A},\,\boldsymbol\beta\right)\boldsymbol\beta}{\beta^2},
\end{equation}
we write down the parallel and perpendicular to the velocity vectors of electric current density:
\begin{equation}
{\textbf{j}}_{\|}=\widehat{\boldsymbol\beta\boldsymbol\beta}\,{\textbf{j}},
\end{equation}
\begin{equation}
{\textbf{j}}_{\perp}=\left(\hat{1}-\widehat{\boldsymbol\beta\boldsymbol\beta}\right){\textbf{j}}.
\end{equation}
The proper plasma frame electric current density after the Lorentz transformation is equal the following expression:
\begin{equation}
{\textbf{j}}^{'}=-c\Gamma\boldsymbol\beta\rho+\left[\hat{1}+(\Gamma-1)\widehat{\boldsymbol\beta\boldsymbol\beta}\right]{\textbf{j}}.
\end{equation}
This expression may be used to calculate the emitting plasma number density proportional the proper frame electric current density squared as suggested by \citet{Lyutikov2005}.

\bsp    
\label{lastpage}

\end{document}